\documentclass{article}
\usepackage[preprint,nonatbib]{neurips_2020}

\usepackage[utf8]{inputenc} 
\usepackage[T1]{fontenc}    
\usepackage{hyperref}       
\usepackage{url}            
\usepackage{booktabs}       
\usepackage{amsfonts}       
\usepackage{nicefrac}       
\usepackage{microtype}      
\usepackage[numbers]{natbib}
\usepackage{textpos}
\usepackage{tabularx,longtable,multirow}
\usepackage[font={small}]{caption}
\usepackage[table]{xcolor}
\usepackage{fncylab} 
\usepackage{fancyhdr} 
\usepackage[english]{babel}
\usepackage{amsmath}
\usepackage{graphicx}
\usepackage{dsfont}
\usepackage{epstopdf} 
\usepackage{backref} 
\usepackage{array}
\usepackage{latexsym}
\usepackage{enumerate}
\usepackage{lscape}
\usepackage{subcaption}
\usepackage[normalem]{ulem}
\usepackage[title]{appendix}

\title{Audio Barlow Twins: Self-Supervised Audio Representation Learning}

\author{%
  Jonah Anton \\
  GLAM, Imperial College London, UK \\
  \texttt{jonahlouisanton@gmail.com} \\
  \And
  Harry Coppock \\
  GLAM, Imperial College London, UK \\
  \texttt{harry.coppock@imperial.ac.uk} \\
  \And
  Pancham Shukla \\
  Imperial College London, UK \\
  \texttt{panchamkumar.shukla@imperial.ac.uk} \\
  \And
  Bj\"{o}rn W. Schuller \\
  GLAM, Imperial College London, UK $\&$ EIHW, University of Augsburg, Germany\\
  \texttt{bjoern.schuller@imperial.ac.uk} \\
}

\begin{document}

\maketitle

\begin{abstract}
The Barlow Twins self-supervised learning objective requires neither negative samples or asymmetric learning updates, achieving results on a par with the current state-of-the-art within Computer Vision. As such, we present \textbf{\textit{Audio Barlow Twins}}, a novel self-supervised audio representation learning approach, adapting Barlow Twins to the audio domain. We pre-train on the large-scale audio dataset AudioSet, and evaluate the quality of the learnt representations on $18$ tasks from the HEAR 2021 Challenge, achieving results which outperform, or otherwise are on a par with, the current state-of-the-art for instance discrimination self-supervised learning approaches to audio representation learning. Code at \url{https://github.com/jonahanton/SSL_audio}.
\end{abstract}

\section{Introduction}

Inspired by recent successes in Computer Vision (CV) \cite{simclr,byol,dino} and Natural Language Processing (NLP) \cite{bert,roberta} in the generation of universal representations\footnote{A representation is a lower-dimensional and compressed, but highly informative, distillation of an input.} through self-supervised learning (SSL) methodologies, much recent interest has been dedicated to using SSL to learn universal representations of audio data \cite{cola, byol-a, ssast, mae-listen}. Whilst generative approaches \cite{ssast, mae-listen} have produced state-of-the-art (SOTA) results for SSL methods in many audio tasks, the current SOTA SSL techniques in CV are dominated by instance discrimination (ID) approaches \cite{ibot,mocov3,msn}, which build a meaningful representation space through training an encoder network to embed similar instances near one another.\par

Barlow Twins \cite{barlow_twins} is one such ID approach, which encourages the empirical cross-correlation matrix between the embeddings of two views of a mini-batch of data samples towards the identity matrix. Through forcing the cross-correlation matrix to the identity, Barlow Twins embeds instances which encode similar semantic content near one another whilst minimising the redundancy between the individual components of the extracted embedding vectors, encouraging the latent representations to be maximally informative. Barlow Twins requires neither negative samples \cite{simclr} nor asymmetric learning updates \cite{byol,simsiam}, instead preventing representational collapse by design. As a result, Barlow Twins i) directly enforces invariances to the applied data augmentations without having to sample negative pairs, and ii) prevents representational collapse in an intuitive and explainable manner \cite{barlow_twins_hsic}, unlike approaches such as BYOL \cite{byol} which are theoretically poorly understood (although some attempts have recently been made \cite{asymm_theory}).  Within the audio domain, the sampling of negative pairs is also potentially problematic, since obtaining such a pair from two different audio signals within a mini-batch \cite{cola,fonseca} can lead to low-quality solutions since two signals may share common sounds, such as a chord sequence in music.\par
It seems reasonable, therefore, that Barlow Twins, when adapted to the audio domain, would produce robust and generalisable audio representations. To this end, we present \textbf{\textit{Audio Barlow Twins}} (ABT), a novel self-supervised audio representation learning method which adapts Barlow Twins \cite{barlow_twins} to the audio domain. ABT achieves results which outperform, or otherwise are on a par with, the current state-of-the-art for ID self-supervised learning approaches to audio representation learning.

\section{Background} 

\textbf{Instance Discrimination}\quad Instance discrimination (ID) SSL approaches \cite{simclr,byol,swav} are built on the core idea of similarity: instances which encode similar semantic content should be embedded near one another in representation space. These methods make use of a Siamese network, where each ‘arm’ of the network processes a different view of the data sample. The extracted feature representations of the two views are then pushed together. Solely enforcing representational similarity of positive pairs is vulnerable to mode collapse onto a constant vector for all inputs, a phenomenon known as representational collapse. Contrastive ID approaches, such as SimCLR \cite{simclr}, prevent representational collapse through the use of negative pairs, which are forced apart in representation space. Non-contrastive ID approaches such as BYOL \cite{byol} prevent representation collapse, instead, through introduction of asymmetry into the learning framework.\par 

\textbf{Audio SSL (ID)}\quad
Many self-supervised learning methods have been proposed to learn generalisable audio representations\footnote{A full and in-depth analysis on the current SOTA audio self-supervised learning methods can be found in the survey produced by \cite{ssl_audio_survey}.}.\citet{fonseca,cola,clar} all adapt SimCLR \cite{simclr} to the audio domain. \citet{fonseca} additionally propose an augmentation which they term \emph{mix-back}, where the incoming spectrogram is mixed with another clip randomly drawn from the training dataset whilst ensuring that the incoming patch remains dominant. \citet{byol-a} present BYOL-A, adapting BYOL \cite{byol} to the audio domain with minimal modifications from the original learning framework. The key modification they make is their proposed data augmentation module, used to generate the two spectrogram views. BYOL-A also makes use of a lightweight convolutional encoder architecture, based on a network used in the solution of the NTT DCASE2020 Challenge Task $6$ (Automated Audio Captioning) \cite{audiontt}, which we use in our experiments and term the \textit{AudioNTT} encoder.

\section{Method}

\begin{figure}[h]
\centering
\includegraphics[width=0.6\textwidth]{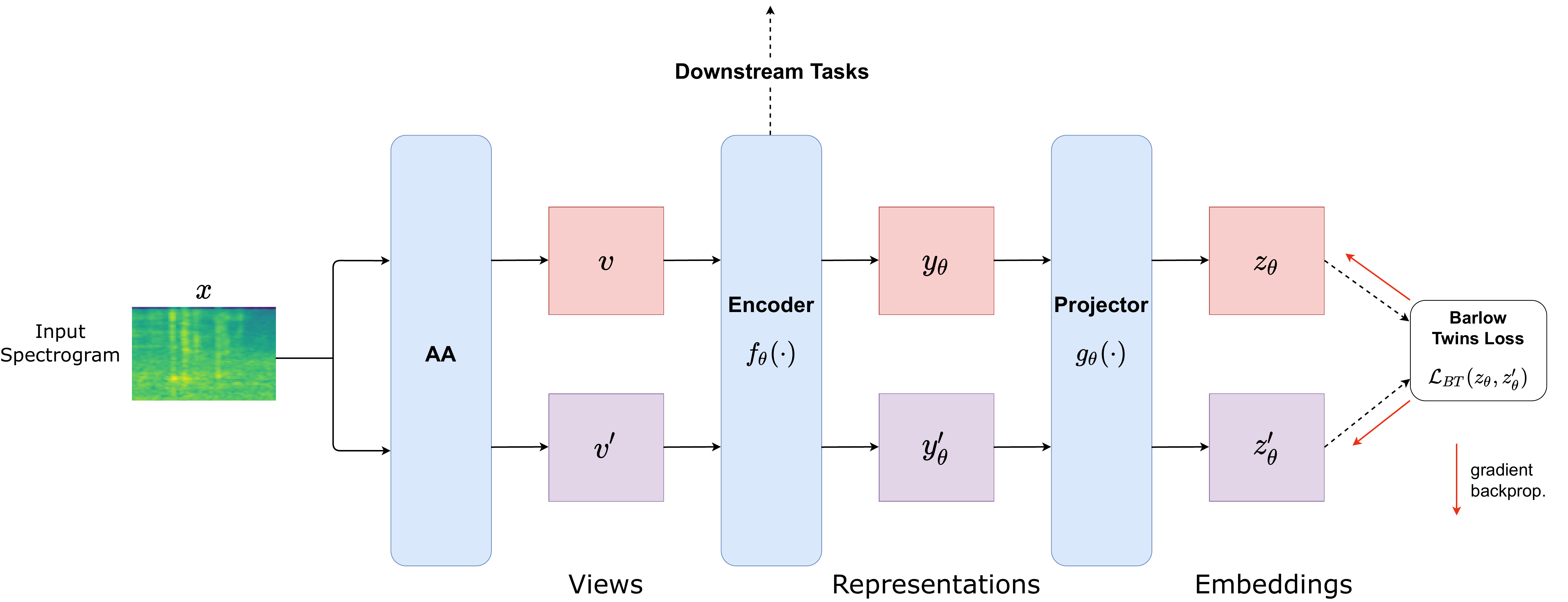}
\caption{
The Audio Barlow Twins learning framework.
}
\label{Audio Barlow Twins Learning Framework Schematic}
\end{figure}

A schematic depicting ABT's high-level architecture is detailed in Figure~\ref{Audio Barlow Twins Learning Framework Schematic}.

\textbf{Generation of views}\quad ABT first produces two views, $v, v^{\prime}$ of an input spectrogram $x$ by stochastic application of the audio augmentation (AA) module $v, v^{\prime} \sim \text{AA}(x)$. The audio augmentation module consists of three different augmentation blocks: Mixup, Random Resize Crop (RRC), and Random Linear Fader (RLF) \cite{byol-av2}. The spectrogram input is first normalised by the dataset mean and standard deviation.\par

\textbf{Extraction of embeddings}\quad The two views are passed through the encoder to obtain the representations, $y_{\theta} = f_{\theta}(v), y_{\theta}^{\prime} = f_{\theta}(v^{\prime})$. The representations are then passed through the projector network to obtain the embeddings, $z_{\theta} = g_{\theta}(y_{\theta}), z_{\theta}^{\prime} = g_{\theta}(y_{\theta}^{\prime})$.\par

\textbf{Barlow Twins objective}\quad The Barlow Twins objective, $\mathcal{L}_{BT}$, is calculated on the embeddings, $\mathcal{L}_{BT}(z_{\theta}, z_{\theta}^{\prime})$. $\mathcal{L}_{BT}$, since it uses batch statistics in its calculation of the embeddings' cross-correlation matrix $C$, cannot in practice be calculated on an input-by-input basis, but instead must be calculated over a batch of embeddings $Z_{\theta}, Z_{\theta}^{\prime}$, with $Z_{\theta} = [z_{\theta}^{1}, ..., z_{\theta}^{B}] \in \mathbb{R}^{B \times d}$, and likewise for $Z_{\theta}^{\prime}$. Formally
\begin{equation}\label{Barlow Twins loss}
\mathcal{L}_{BT} = \alpha \sum_{i} (1 - C_{ii})^{2} + \lambda \sum_{i\neq j} C_{ij}^{2},
\end{equation}
where the first term enforces representational invariance to the applied audio augmentations, and the second term minimises the redundancy between the individual components of the embedding vectors. The positive constants $\alpha$ and $\lambda$ control the trade-off between the importance of these two terms, and by default $\alpha$ is set to $1$ and $\lambda$ to $0.005$ (as in the original publication \cite{barlow_twins}). The cross-correlation matrix $C$ is computed between the embeddings within the batch $B$,
\begin{equation}\label{cross-correlation matrix}
C_{ij} = \sum_{b=1}^{B} \hat{Z}_{\theta, i}^{b} \hat{Z^{\prime}}_{\theta, j}^{b},
\end{equation}
where $\hat{Z}_{\theta}$ is the normalised embedding $Z_{\theta}$ along the batch dimension, and $\hat{Z}_{\theta, i}^{b}$ corresponds to the $i^{th}$ component of the $b^{th}$ batch element of $\hat{Z}_{\theta}$.

\section{Experiments}

We pre-train on the large-scale audio dataset AudioSet \cite{audioset} for $100$ epochs with a batch size of $128$, which corresponds to $\sim 1.3$M training iterations. We successfully download $1,629,756$ clips (corresponding to $\sim 4,500$ hours of audio) from AudioSet's unbalanced train subset, which are used for ABT pre-training.\par

\textbf{Audio preprocessing}\quad  All audio samples are converted with a sampling frequency of $16$ kHz to (log-scaled) mel-spectrograms using a $64$\,ms sliding window with a $10$ ms step size, extracting $F = 64$ mel frequency bins in the range $60-7,800$\,Hz. By default, during pre-training, we randomly crop $T = 96$ time frames (all clips with shorter duration are padded with zeros), corresponding to $950$\,ms of audio. This produces a mel-spectrogram of size $F \times T = 64 \times 96$.\par

\textbf{Architecture}\quad We consider two encoders, the AudioNTT convolutional encoder \cite{audiontt}, and the ViT$_{C}$ encoder \cite{vitc}. We consider the ViT$_{C}$-\textit{B}(ase)\footnote{The ViT$_{C}$-\textit{B} corresponds to the ViT$_{C}$-18GF model proposed in the original publication \cite{vitc}.} model, using a patch size of $16 \times 8$. A learnable \texttt{[CLS]} token is prepended to the sequence of patches, and its output representation, $O_{\texttt{[CLS]}}$, is taken as representative of the clip as a whole. Fixed sinusoidal positional encodings are added to each patch.\par

\textbf{Downstream Tasks}\quad We use $18$ tasks from the HEAR 2021 Challenge \cite{hear} for evaluation. HEAR includes two types of tasks, i) scene-based tasks, corresponding to classification of an entire audio clip, and ii) timestamp-based tasks, corresponding to sound event detection and or transcription over time. For each task, the representations are extracted from the frozen pre-trained model and then evaluated using the \texttt{hear-eval}\footnote{\url{https://github.com/hearbenchmark/hear-eval-kit}} toolkit, which trains a shallow Multilayer Perceptron (MLP) classifier on the extracted representations. 

\section{Results}\label{sec:results}

We compare the performance of the ABT pre-trained models on the $18$ HEAR tasks with two baseline models, CREPE \cite{crepe}, wav2vec2.0 \cite{wav2vec2.0}, and to BYOL-A$^{\ast}$\footnote{BYOL-A$^{\ast}$ is a reimplementation of BYOL-A \cite{byol-a} by \citet{byol-s}, which we use since \citet{byol-a} did not evaluate on the HEAR tasks.} \cite{byol-a}. The results for the scene-based and timestamp-based tasks are detailed in Tables~\ref{tab:results-HEAR-AudioSet-speech+env},\ref{tab:results-HEAR-AudioSet-music} and \ref{tab:results-HEAR-AudioSet-timestamp}.\par 

ABT, with the AudioNTT encoder, generally performs on a par with, or outperforms, BYOL-A$^{\ast}$, which uses the same AudioNTT encoder architecture, on the scene-based tasks, and consistently outperforms BYOL-A$^{\ast}$ on the timestamp-based tasks. We see further consistent improvements over the HEAR baseline models (CREPE, wav2vec2.0), except on the type of tasks on which these models have been specialised (music for CREPE, speech for wav2vec2.0). These results demonstrate the robustness of ABT in the generation of general-purpose audio representations. Interestingly, ABT pre-training appears damaging to performance on several of the music tasks, often leading to performance degradation from the random baselines. We find this to be particularly evident for the Mridingam Stroke and Tonic, NSynth ($5$h and $50$h), and MAESTRO tasks. This extends to other ID methods, with BYOL-A$^{\ast}$, performing similarly poorly. The aforementioned tasks all require a sound's pitch to be correctly discerned. However, invariance to pitch perturbations is enforced through RRC, and as such, it is intuitive that a model will consequently struggle to classify pitch. That said, we find through extensive ablation studies, detailed in Appendix~\ref{sec:ablat}, that RRC does considerably improve the quality of the learnt representations. We therefore observe an issue with transferring ID methods from CV to audio, since such methods rely on applying data augmentations to generate two views, and any given data augmentation may benefit one type of audio task but harm another. This provides support for generative self-supervised methods for learning universal audio representations \cite{ssast, mae-listen}, since they don't require the use of any data augmentations.

\begin{table}[t!] \centering
\caption{
Results on HEAR speech and environmental sound scene-based tasks. Top two performing models for each task are shown \underline{underlined} and highlighted. $\% \uparrow_{\text{RAND}}$ refers to the fractional increase (\textbf{not} the absolute increase) from the average score obtained by the random baseline for that model.
}
\label{tab:results-HEAR-AudioSet-speech+env}
\resizebox{\textwidth}{!}{%
\begin{tabular}{l c c c c c c c c c c c c c c}
\hline
& \multicolumn{8}{c}{\textbf{Speech}} & & \multicolumn{5}{c}{\textbf{Environmental Sound}} \\
\cline{2-9}\cline{11-15}
\textbf{Model} & CREMA-D & LbC & SPC-$5$h & SPC-F & VocIm & VoxL & & \textbf{Avg} ($\% \uparrow_{\text{RAND}}$) & & ESC-50 & FSD50K & Gunshot & & \textbf{Avg} ($\% \uparrow_{\text{RAND}}$) \\
\hline
$[\text{HEAR}]$ CREPE & $0.383$ & $0.499$ & $0.180$ & $0.211$ & $0.051$ & $0.142$ & & $0.244$ & & $0.301$ & $0.159$ & $0.863$ & & $0.441$ \\
$[\text{HEAR}]$ wav2vec2.0 & \cellcolor{blue!25} \underline{$0.656$} & $0.692$ & $0.838$ & $0.879$ & $0.080$ & \cellcolor{blue!25} \underline{$0.493$} & & \cellcolor{blue!25} \underline{$0.606$} & & $0.561$ & $0.342$ & $0.848$ & & $0.584$ \\
$[\text{HEAR}]$ BYOL-A$^{\ast}$ & \cellcolor{blue!25} \underline{$0.623$} & \cellcolor{blue!25} \underline{$0.788$} & \cellcolor{blue!25} \underline{$0.896$} & \cellcolor{blue!25} \underline{$0.924$} & \cellcolor{blue!25} \underline{$0.137$} & \cellcolor{blue!25} \underline{$0.390$} & & \cellcolor{blue!25} \underline{$0.626$} & & \cellcolor{blue!25} \underline{$0.789$} & \cellcolor{blue!25} \underline{$0.489$} & \cellcolor{blue!25} \underline{$0.875$} & & \cellcolor{blue!25} \underline{$0.7180$}\\
\hline
$[\text{ABT}]$ AudioNTT & $0.594$ & $0.745$ & \cellcolor{blue!25} \underline{$0.882$} & \cellcolor{blue!25} \underline{$0.910$} & \cellcolor{blue!25} \underline{$0.111$} & $0.324$ & & $0.594$ ($17\%$) & & \cellcolor{blue!25} \underline{$0.786$} & \cellcolor{blue!25} \underline{$0.474$} & \cellcolor{blue!25} \underline{$0.905$} & & \cellcolor{blue!25} \underline{$0.721$} ($24\%$)\\
$[\text{ABT}]$ ViT$_{C}$\textit{-B} ($16 \times 8$) & $0.581$ & \cellcolor{blue!25} \underline{$0.812$} & $0.724$ & $0.771$ & $0.087$ & $0.312$ & & $0.548$ ($140\%$) & & $0.705$ & $0.446$ & $0.845$ & & $0.666$ ($49\%$) \\
\hline
\end{tabular}}
\end{table}

\begin{table}[t!] \centering
\caption{
Results on HEAR music scene-based tasks.
}
\label{tab:results-HEAR-AudioSet-music}
\resizebox{\textwidth}{!}{%
\begin{tabular}{l c c c c c c c c c}
\hline
& \multicolumn{9}{c}{\textbf{Music}} \\
\cline{2-10}
\textbf{Model} & Beijing & GTZAN-Genre & GTZAN-M/S & Mrd-Stroke & Mrd-Tonic & NSynth $5$h & NSynth $50$h & & \textbf{Avg} ($\% \uparrow_{\text{RAND}}$) \\
\hline
$[\text{HEAR}]$ CREPE & \cellcolor{blue!25} \underline{$0.928$} & $0.645$ & $0.929$ & $0.898$ & $0.824$ & \cellcolor{blue!25} \underline{$0.870$} & \cellcolor{blue!25} \underline{$0.900$} & & \cellcolor{blue!25} \underline{$0.856$} \\
$[\text{HEAR}]$ wav2vec2.0 & $0.907$ & $0.780$ & $0.946$ & $0.943$ & $0.828$ & $0.402$ & $0.653$ & & $0.780$ \\
$[\text{HEAR}]$ BYOL-A$^{\ast}$ & $0.919$ & \cellcolor{blue!25} \underline{$0.835$} & \cellcolor{blue!25} \underline{$0.969$} & \cellcolor{blue!25} \underline{$0.970$} & \cellcolor{blue!25} \underline{$0.900$} & $0.290$ & $0.642$ & & $0.789$\\
\hline
$[\text{ABT}]$ AudioNTT & \cellcolor{blue!25} \underline{$0.966$} & \cellcolor{blue!25} \underline{$0.818$} & $0.962$ & \cellcolor{blue!25} \underline{$0.970$} & \cellcolor{blue!25} \underline{$0.932$} & \cellcolor{blue!25} \underline{$0.476$} & \cellcolor{blue!25} \underline{$0.740$} & & \cellcolor{blue!25} \underline{$0.838$} ($-1\%$) \\
$[\text{ABT}]$ ViT$_{C}$\textit{-B} ($16 \times 8$) & $0.869$ & $0.765$ & \cellcolor{blue!25} \underline{$0.992$} & $0.952$ & $0.897$ & $0.280$ & $0.632$ & & $0.769$ ($15\%$) \\
\hline
\end{tabular}}
\end{table}

\begin{table}[t!] \centering
\caption{
Results on HEAR timestamp-based tasks.Error rate ($\downarrow$) indicates that a lower error rate is better. Table format adapted from \cite{byol-s}.
}
\label{tab:results-HEAR-AudioSet-timestamp}
\resizebox{0.8\textwidth}{!}{%
\begin{tabular}{l c c c c c c}
\hline
& \multicolumn{2}{c}{\textbf{DCASE}} &  & \multicolumn{2}{c}{\textbf{MAESTRO}} & \multicolumn{1}{c}{\textbf{Avg}} ($\% \uparrow_{\text{RAND}}$) \\
\cline{2-3}\cline{5-6}
\textbf{Model} & Onset FMS & Error rate ($\downarrow$) & & Onset FMS & Onset w/ Offset FMS & Onset FMS \\
\hline
$[\text{HEAR}]$ CREPE & $0.552$ & $0.420$ & & \cellcolor{blue!25} \underline{$0.3910$} & \cellcolor{blue!25} \underline{$0.15$} & \cellcolor{blue!25} \underline{$0.472$} \\
$[\text{HEAR}]$ wav2vec2.0 & $0.670$ & $0.320$ & & $0.0328$ & $0.009$ & $0.351$ \\
$[\text{HEAR}]$ BYOL-A$^{\ast}$ & $0.499$ & $0.503$ & & $0.0028$ & $0.00029$ & $0.251$ \\
\hline
$[\text{ABT}]$ AudioNTT & \cellcolor{blue!25} \underline{$0.761$} & \cellcolor{blue!25} \underline{$0.274$} & & \cellcolor{blue!25} \underline{$0.04801$} & \cellcolor{blue!25} \underline{$0.00672$} & \cellcolor{blue!25} \underline{$0.405$} ($27\%$) \\
$[\text{ABT}]$ ViT$_{C}$\textit{-B} ($16 \times 8$) & \cellcolor{blue!25} \underline{$0.722$} & \cellcolor{blue!25} \underline{$0.275$} & & $0.0263$ & $0.00429$ & $0.374$ ($-10\%$) \\
\hline
\end{tabular}}
\end{table}

\section{Conclusion}

In this paper, we presented \textit{\textbf{Audio Barlow Twins}} (ABT), a novel self-supervised audio representation learning method which adapts Barlow Twins \cite{barlow_twins} to the audio domain. ABT pre-training on AudioSet \cite{audioset} for $100$ epochs with the AudioNTT encoder \cite{audiontt} results in model performance which is on a par with, and in several cases better than, BYOL-A \cite{byol-a}. We found commonly introduced augmentations to be harmful to ABT in certain settings. Future works should consider the effect on different downstream tasks of different augmentations that act directly on raw waveforms, within the ABT learning framework. Applying the augmentations directly on raw waveforms, instead of spectrograms, allows for a) a better control of the strength of these augmentations (as it is possible to \textit{listen} directly to their effect), and b) a greater number of augmentations to be considered (e.g. pitch shift, time masking, time shift, time stretch, fade in/out, compression, etc.). We were unable to apply data augmentations during training directly on raw waveforms in this work due to an I/O bottleneck.

\newpage
\bibliographystyle{IEEEtranN}
\bibliography{references.bib}

\begin{thebibliography}{37}
\providecommand{\natexlab}[1]{#1}
\providecommand{\url}[1]{#1}
\csname url@samestyle\endcsname
\providecommand{\newblock}{\relax}
\providecommand{\bibinfo}[2]{#2}
\providecommand{\BIBentrySTDinterwordspacing}{\spaceskip=0pt\relax}
\providecommand{\BIBentryALTinterwordstretchfactor}{4}
\providecommand{\BIBentryALTinterwordspacing}{\spaceskip=\fontdimen2\font plus
\BIBentryALTinterwordstretchfactor\fontdimen3\font minus
  \fontdimen4\font\relax}
\providecommand{\BIBforeignlanguage}[2]{{%
\expandafter\ifx\csname l@#1\endcsname\relax
\typeout{** WARNING: IEEEtranN.bst: No hyphenation pattern has been}%
\typeout{** loaded for the language `#1'. Using the pattern for}%
\typeout{** the default language instead.}%
\else
\language=\csname l@#1\endcsname
\fi
#2}}
\providecommand{\BIBdecl}{\relax}
\BIBdecl

\bibitem[Chen et~al.(2020)Chen, Kornblith, Norouzi, and Hinton]{simclr}
\BIBentryALTinterwordspacing
T.~Chen, S.~Kornblith, M.~Norouzi, and G.~E. Hinton, ``A simple framework for
  contrastive learning of visual representations,'' \emph{CoRR}, vol.
  abs/2002.05709, 2020. [Online]. Available:
  \url{https://arxiv.org/abs/2002.05709}
\BIBentrySTDinterwordspacing

\bibitem[Grill et~al.(2020)Grill, Strub, Altch{\'{e}}, Tallec, Richemond,
  Buchatskaya, Doersch, Pires, Guo, Azar, Piot, Kavukcuoglu, Munos, and
  Valko]{byol}
\BIBentryALTinterwordspacing
J.~Grill, F.~Strub, F.~Altch{\'{e}}, C.~Tallec, P.~H. Richemond,
  E.~Buchatskaya, C.~Doersch, B.~{\'{A}}. Pires, Z.~D. Guo, M.~G. Azar,
  B.~Piot, K.~Kavukcuoglu, R.~Munos, and M.~Valko, ``Bootstrap your own latent:
  {A} new approach to self-supervised learning,'' \emph{CoRR}, vol.
  abs/2006.07733, 2020. [Online]. Available:
  \url{https://arxiv.org/abs/2006.07733}
\BIBentrySTDinterwordspacing

\bibitem[Caron et~al.(2021)Caron, Touvron, Misra, J{\'{e}}gou, Mairal,
  Bojanowski, and Joulin]{dino}
\BIBentryALTinterwordspacing
M.~Caron, H.~Touvron, I.~Misra, H.~J{\'{e}}gou, J.~Mairal, P.~Bojanowski, and
  A.~Joulin, ``Emerging properties in self-supervised vision transformers,''
  \emph{CoRR}, vol. abs/2104.14294, 2021. [Online]. Available:
  \url{https://arxiv.org/abs/2104.14294}
\BIBentrySTDinterwordspacing

\bibitem[Devlin et~al.(2018)Devlin, Chang, Lee, and Toutanova]{bert}
\BIBentryALTinterwordspacing
J.~Devlin, M.~Chang, K.~Lee, and K.~Toutanova, ``{BERT:} pre-training of deep
  bidirectional transformers for language understanding,'' \emph{CoRR}, vol.
  abs/1810.04805, 2018. [Online]. Available:
  \url{http://arxiv.org/abs/1810.04805}
\BIBentrySTDinterwordspacing

\bibitem[Liu et~al.(2019)Liu, Ott, Goyal, Du, Joshi, Chen, Levy, Lewis,
  Zettlemoyer, and Stoyanov]{roberta}
\BIBentryALTinterwordspacing
Y.~Liu, M.~Ott, N.~Goyal, J.~Du, M.~Joshi, D.~Chen, O.~Levy, M.~Lewis,
  L.~Zettlemoyer, and V.~Stoyanov, ``Roberta: {A} robustly optimized {BERT}
  pretraining approach,'' \emph{CoRR}, vol. abs/1907.11692, 2019. [Online].
  Available: \url{http://arxiv.org/abs/1907.11692}
\BIBentrySTDinterwordspacing

\bibitem[Saeed et~al.(2020)Saeed, Grangier, and Zeghidour]{cola}
\BIBentryALTinterwordspacing
A.~Saeed, D.~Grangier, and N.~Zeghidour, ``Contrastive learning of
  general-purpose audio representations,'' \emph{CoRR}, vol. abs/2010.10915,
  2020. [Online]. Available: \url{https://arxiv.org/abs/2010.10915}
\BIBentrySTDinterwordspacing

\bibitem[Niizumi et~al.(2021)Niizumi, Takeuchi, Ohishi, Harada, and
  Kashino]{byol-a}
\BIBentryALTinterwordspacing
D.~Niizumi, D.~Takeuchi, Y.~Ohishi, N.~Harada, and K.~Kashino, ``Byol for
  audio: Self-supervised learning for general-purpose audio representation,''
  2021. [Online]. Available: \url{https://arxiv.org/abs/2103.06695}
\BIBentrySTDinterwordspacing

\bibitem[Gong et~al.(2021)Gong, Lai, Chung, and Glass]{ssast}
\BIBentryALTinterwordspacing
Y.~Gong, C.~J. Lai, Y.~Chung, and J.~R. Glass, ``{SSAST:} self-supervised audio
  spectrogram transformer,'' \emph{CoRR}, vol. abs/2110.09784, 2021. [Online].
  Available: \url{https://arxiv.org/abs/2110.09784}
\BIBentrySTDinterwordspacing

\bibitem[Huang et~al.(2022)Huang, Xu, Li, Baevski, Auli, Galuba, Metze, and
  Feichtenhofer]{mae-listen}
\BIBentryALTinterwordspacing
P.-Y. Huang, H.~Xu, J.~Li, A.~Baevski, M.~Auli, W.~Galuba, F.~Metze, and
  C.~Feichtenhofer, ``Masked autoencoders that listen,'' 2022. [Online].
  Available: \url{https://arxiv.org/abs/2207.06405}
\BIBentrySTDinterwordspacing

\bibitem[Zhou et~al.(2021)Zhou, Wei, Wang, Shen, Xie, Yuille, and Kong]{ibot}
\BIBentryALTinterwordspacing
J.~Zhou, C.~Wei, H.~Wang, W.~Shen, C.~Xie, A.~L. Yuille, and T.~Kong, ``ibot:
  Image {BERT} pre-training with online tokenizer,'' \emph{CoRR}, vol.
  abs/2111.07832, 2021. [Online]. Available:
  \url{https://arxiv.org/abs/2111.07832}
\BIBentrySTDinterwordspacing

\bibitem[Chen et~al.(2021)Chen, Xie, and He]{mocov3}
\BIBentryALTinterwordspacing
X.~Chen, S.~Xie, and K.~He, ``An empirical study of training self-supervised
  vision transformers,'' \emph{CoRR}, vol. abs/2104.02057, 2021. [Online].
  Available: \url{https://arxiv.org/abs/2104.02057}
\BIBentrySTDinterwordspacing

\bibitem[Assran et~al.(2022)Assran, Caron, Misra, Bojanowski, Bordes, Vincent,
  Joulin, Rabbat, and Ballas]{msn}
\BIBentryALTinterwordspacing
M.~Assran, M.~Caron, I.~Misra, P.~Bojanowski, F.~Bordes, P.~Vincent, A.~Joulin,
  M.~Rabbat, and N.~Ballas, ``Masked siamese networks for label-efficient
  learning,'' 2022. [Online]. Available: \url{https://arxiv.org/abs/2204.07141}
\BIBentrySTDinterwordspacing

\bibitem[Zbontar et~al.(2021)Zbontar, Jing, Misra, LeCun, and
  Deny]{barlow_twins}
\BIBentryALTinterwordspacing
J.~Zbontar, L.~Jing, I.~Misra, Y.~LeCun, and S.~Deny, ``Barlow twins:
  Self-supervised learning via redundancy reduction,'' \emph{CoRR}, vol.
  abs/2103.03230, 2021. [Online]. Available:
  \url{https://arxiv.org/abs/2103.03230}
\BIBentrySTDinterwordspacing

\bibitem[Chen and He(2020)]{simsiam}
\BIBentryALTinterwordspacing
X.~Chen and K.~He, ``Exploring simple siamese representation learning,''
  \emph{CoRR}, vol. abs/2011.10566, 2020. [Online]. Available:
  \url{https://arxiv.org/abs/2011.10566}
\BIBentrySTDinterwordspacing

\bibitem[Tsai et~al.(2021)Tsai, Bai, Morency, and
  Salakhutdinov]{barlow_twins_hsic}
\BIBentryALTinterwordspacing
Y.-H.~H. Tsai, S.~Bai, L.-P. Morency, and R.~Salakhutdinov, ``A note on
  connecting barlow twins with negative-sample-free contrastive learning,''
  2021. [Online]. Available: \url{https://arxiv.org/abs/2104.13712}
\BIBentrySTDinterwordspacing

\bibitem[Tian et~al.(2021)Tian, Chen, and Ganguli]{asymm_theory}
\BIBentryALTinterwordspacing
Y.~Tian, X.~Chen, and S.~Ganguli, ``Understanding self-supervised learning
  dynamics without contrastive pairs,'' \emph{CoRR}, vol. abs/2102.06810, 2021.
  [Online]. Available: \url{https://arxiv.org/abs/2102.06810}
\BIBentrySTDinterwordspacing

\bibitem[Fonseca et~al.(2020{\natexlab{a}})Fonseca, Ortego, McGuinness,
  O'Connor, and Serra]{fonseca}
\BIBentryALTinterwordspacing
E.~Fonseca, D.~Ortego, K.~McGuinness, N.~E. O'Connor, and X.~Serra,
  ``Unsupervised contrastive learning of sound event representations,'' 2020.
  [Online]. Available: \url{https://arxiv.org/abs/2011.07616}
\BIBentrySTDinterwordspacing

\bibitem[Caron et~al.(2020)Caron, Misra, Mairal, Goyal, Bojanowski, and
  Joulin]{swav}
\BIBentryALTinterwordspacing
M.~Caron, I.~Misra, J.~Mairal, P.~Goyal, P.~Bojanowski, and A.~Joulin,
  ``Unsupervised learning of visual features by contrasting cluster
  assignments,'' \emph{CoRR}, vol. abs/2006.09882, 2020. [Online]. Available:
  \url{https://arxiv.org/abs/2006.09882}
\BIBentrySTDinterwordspacing

\bibitem[Liu et~al.(2022)Liu, Mallol-Ragolta, Parada-Cabeleiro, Qian, Jing,
  Kathan, Hu, and Schuller]{ssl_audio_survey}
\BIBentryALTinterwordspacing
S.~Liu, A.~Mallol-Ragolta, E.~Parada-Cabeleiro, K.~Qian, X.~Jing, A.~Kathan,
  B.~Hu, and B.~W. Schuller, ``Audio self-supervised learning: A survey,''
  2022. [Online]. Available: \url{https://arxiv.org/abs/2203.01205}
\BIBentrySTDinterwordspacing

\bibitem[Al-Tahan and Mohsenzadeh(2020)]{clar}
\BIBentryALTinterwordspacing
H.~Al-Tahan and Y.~Mohsenzadeh, ``Clar: Contrastive learning of auditory
  representations,'' 2020. [Online]. Available:
  \url{https://arxiv.org/abs/2010.09542}
\BIBentrySTDinterwordspacing

\bibitem[Koizumi et~al.(2020)Koizumi, Takeuchi, Ohishi, Harada, and
  Kashino]{audiontt}
\BIBentryALTinterwordspacing
Y.~Koizumi, D.~Takeuchi, Y.~Ohishi, N.~Harada, and K.~Kashino, ``The ntt
  dcase2020 challenge task 6 system: Automated audio captioning with keywords
  and sentence length estimation,'' 2020. [Online]. Available:
  \url{https://arxiv.org/abs/2007.00225}
\BIBentrySTDinterwordspacing

\bibitem[Niizumi et~al.(2022)Niizumi, Takeuchi, Ohishi, Harada, and
  Kashino]{byol-av2}
\BIBentryALTinterwordspacing
D.~Niizumi, D.~Takeuchi, Y.~Ohishi, N.~Harada, and K.~Kashino, ``Byol for
  audio: Exploring pre-trained general-purpose audio representations,'' 2022.
  [Online]. Available: \url{https://arxiv.org/abs/2204.07402}
\BIBentrySTDinterwordspacing

\bibitem[Gemmeke et~al.(2017)Gemmeke, Ellis, Freedman, Jansen, Lawrence, Moore,
  Plakal, and Ritter]{audioset}
J.~F. Gemmeke, D.~P.~W. Ellis, D.~Freedman, A.~Jansen, W.~Lawrence, R.~C.
  Moore, M.~Plakal, and M.~Ritter, ``Audio set: An ontology and human-labeled
  dataset for audio events,'' in \emph{2017 IEEE International Conference on
  Acoustics, Speech and Signal Processing (ICASSP)}, 2017, pp. 776--780.

\bibitem[Xiao et~al.(2021)Xiao, Singh, Mintun, Darrell, Doll{\'{a}}r, and
  Girshick]{vitc}
\BIBentryALTinterwordspacing
T.~Xiao, M.~Singh, E.~Mintun, T.~Darrell, P.~Doll{\'{a}}r, and R.~B. Girshick,
  ``Early convolutions help transformers see better,'' \emph{CoRR}, vol.
  abs/2106.14881, 2021. [Online]. Available:
  \url{https://arxiv.org/abs/2106.14881}
\BIBentrySTDinterwordspacing

\bibitem[Turian et~al.(2022)Turian, Shier, Khan, Raj, Schuller, Steinmetz,
  Malloy, Tzanetakis, Velarde, McNally, Henry, Pinto, Noufi, Clough, Herremans,
  Fonseca, Engel, Salamon, Esling, Manocha, Watanabe, Jin, and Bisk]{hear}
\BIBentryALTinterwordspacing
J.~Turian, J.~Shier, H.~R. Khan, B.~Raj, B.~W. Schuller, C.~J. Steinmetz,
  C.~Malloy, G.~Tzanetakis, G.~Velarde, K.~McNally, M.~Henry, N.~Pinto,
  C.~Noufi, C.~Clough, D.~Herremans, E.~Fonseca, J.~Engel, J.~Salamon,
  P.~Esling, P.~Manocha, S.~Watanabe, Z.~Jin, and Y.~Bisk, ``Hear: Holistic
  evaluation of audio representations,'' 2022. [Online]. Available:
  \url{https://arxiv.org/abs/2203.03022}
\BIBentrySTDinterwordspacing

\bibitem[Kim et~al.(2018)Kim, Salamon, Li, and Bello]{crepe}
\BIBentryALTinterwordspacing
J.~W. Kim, J.~Salamon, P.~Li, and J.~P. Bello, ``Crepe: A convolutional
  representation for pitch estimation,'' 2018. [Online]. Available:
  \url{https://arxiv.org/abs/1802.06182}
\BIBentrySTDinterwordspacing

\bibitem[Baevski et~al.(2020)Baevski, Zhou, Mohamed, and Auli]{wav2vec2.0}
\BIBentryALTinterwordspacing
A.~Baevski, H.~Zhou, A.~Mohamed, and M.~Auli, ``wav2vec 2.0: {A} framework for
  self-supervised learning of speech representations,'' \emph{CoRR}, vol.
  abs/2006.11477, 2020. [Online]. Available:
  \url{https://arxiv.org/abs/2006.11477}
\BIBentrySTDinterwordspacing

\bibitem[Elbanna et~al.(2022)Elbanna, Scheidwasser-Clow, Kegler, Beckmann,
  Hajal, and Cernak]{byol-s}
\BIBentryALTinterwordspacing
G.~Elbanna, N.~Scheidwasser-Clow, M.~Kegler, P.~Beckmann, K.~E. Hajal, and
  M.~Cernak, ``Byol-s: Learning self-supervised speech representations by
  bootstrapping,'' 2022. [Online]. Available:
  \url{https://arxiv.org/abs/2206.12038}
\BIBentrySTDinterwordspacing

\bibitem[You et~al.(2017)You, Gitman, and Ginsburg]{lars}
\BIBentryALTinterwordspacing
Y.~You, I.~Gitman, and B.~Ginsburg, ``Large batch training of convolutional
  networks,'' 2017. [Online]. Available: \url{https://arxiv.org/abs/1708.03888}
\BIBentrySTDinterwordspacing

\bibitem[Loshchilov and Hutter(2017)]{adamw}
\BIBentryALTinterwordspacing
I.~Loshchilov and F.~Hutter, ``Fixing weight decay regularization in adam,''
  \emph{CoRR}, vol. abs/1711.05101, 2017. [Online]. Available:
  \url{http://arxiv.org/abs/1711.05101}
\BIBentrySTDinterwordspacing

\bibitem[McFee et~al.(2015)McFee, Raffel, Liang, Ellis, Mcvicar, Battenberg,
  and Nieto]{librosa}
B.~McFee, C.~Raffel, D.~Liang, D.~Ellis, M.~Mcvicar, E.~Battenberg, and
  O.~Nieto, ``librosa: Audio and music signal analysis in python,'' 01 2015,
  pp. 18--24.

\bibitem[Kingma and Ba(2014)]{adam}
\BIBentryALTinterwordspacing
D.~P. Kingma and J.~Ba, ``Adam: A method for stochastic optimization,'' 2014.
  [Online]. Available: \url{https://arxiv.org/abs/1412.6980}
\BIBentrySTDinterwordspacing

\bibitem[Fonseca et~al.(2020{\natexlab{b}})Fonseca, Favory, Pons, Font, and
  Serra]{fsd50k}
\BIBentryALTinterwordspacing
E.~Fonseca, X.~Favory, J.~Pons, F.~Font, and X.~Serra, ``Fsd50k,'' Oct. 2020.
  [Online]. Available: \url{https://doi.org/10.5281/zenodo.4060432}
\BIBentrySTDinterwordspacing

\bibitem[Akiba et~al.(2019)Akiba, Sano, Yanase, Ohta, and Koyama]{optuna}
\BIBentryALTinterwordspacing
T.~Akiba, S.~Sano, T.~Yanase, T.~Ohta, and M.~Koyama, ``Optuna: {A}
  next-generation hyperparameter optimization framework,'' \emph{CoRR}, vol.
  abs/1907.10902, 2019. [Online]. Available:
  \url{http://arxiv.org/abs/1907.10902}
\BIBentrySTDinterwordspacing

\bibitem[Hutter et~al.(2014)Hutter, Hoos, and Leyton-Brown]{fanova}
F.~Hutter, H.~Hoos, and K.~Leyton-Brown, ``An efficient approach for assessing
  hyperparameter importance,'' \emph{31st International Conference on Machine
  Learning, ICML 2014}, vol.~2, pp. 1130--1144, 01 2014.

\bibitem[Dosovitskiy et~al.(2020)Dosovitskiy, Beyer, Kolesnikov, Weissenborn,
  Zhai, Unterthiner, Dehghani, Minderer, Heigold, Gelly, Uszkoreit, and
  Houlsby]{vit}
\BIBentryALTinterwordspacing
A.~Dosovitskiy, L.~Beyer, A.~Kolesnikov, D.~Weissenborn, X.~Zhai,
  T.~Unterthiner, M.~Dehghani, M.~Minderer, G.~Heigold, S.~Gelly, J.~Uszkoreit,
  and N.~Houlsby, ``An image is worth 16x16 words: Transformers for image
  recognition at scale,'' \emph{CoRR}, vol. abs/2010.11929, 2020. [Online].
  Available: \url{https://arxiv.org/abs/2010.11929}
\BIBentrySTDinterwordspacing

\bibitem[He et~al.(2021)He, Chen, Xie, Li, Dollár, and Girshick]{mae}
\BIBentryALTinterwordspacing
K.~He, X.~Chen, S.~Xie, Y.~Li, P.~Dollár, and R.~Girshick, ``Masked
  autoencoders are scalable vision learners,'' 2021. [Online]. Available:
  \url{https://arxiv.org/abs/2111.06377}
\BIBentrySTDinterwordspacing

\end{thebibliography}

\newpage
\begin{appendices}
\renewcommand{\thefigure}{A-\arabic{figure}}
\setcounter{figure}{0}
\renewcommand{\thetable}{A-\arabic{table}}
\setcounter{table}{0}

\section{Further Implementations Details}\label{sec:imp_details}
In this appendix we provide further details for Audio Barlow Twins pre-training and HEAR evaluation.
\subsection{Pre-training}
We pre-train on the unbalanced train subset of AudioSet ($\sim 1.6$M audio segments) for $100$ epochs with a batch size of $128$, which corresponds to $\sim 1.3$M training iterations. We use a smaller version of the projector network than that proposed in the original Barlow Twins publication \cite{barlow_twins}, although with the same modular structure. The projector network corresponds to a small MLP with one hidden layer, which has hidden dimension $8192$, and an output dimension of $1048$. The first layer of the projector is followed by a batch normalisation and Rectified Linear Unit (ReLU) non-linearity. The Barlow Twins loss hyperparameters are set to $1$ and $5 \times 10^{-3}$, for $\alpha$ and $\lambda$ respectively. For the AudioNTT encoder, we use the Layer-wise Adaptive Rate Scaling (LARS) optimizer \cite{lars}, with a learning rate of $0.4$ for the weights and $0.0048$ for the biases and batch normalisation parameters. We use a weight decay of $1 \cdot 10^{-5}$. Following from \cite{barlow_twins}, LARS adaptation, as well as weight decay, do not apply to the biases and batch normalisation parameters. The choice of the LARS optimizer, over Adam and Stochastic gradient descent (SGD), as well as the optimizer hyperparameters (learning rate weights, learning rate biases, weight decay) are selected after conducting an extensive hyperparameter sweep (Appendix \ref{sec:hyps}). We also consider the default values\footnote{Appropriately scaled for a batch size of $128$ (using linear scaling), \cite{barlow_twins} use a learning rate of $0.1$ for the weights and $0.0024$ for the biases, with a weight decay of $1.5 \cdot 10^{-6}$.} as used by \cite{barlow_twins}, but find a noticeable degradation in model performance. For the ViT$_{C}$-\textit{B} encoder, we use AdamW \cite{adamw} with the default hyperparameter values as suggested\footnote{\citet{vitc} perform an extensive hyperparameter sweep for the ViT$_{C}$ with a batch size of $2048$, using a patch size of $16 \times 16$. They find, with AdamW, a \textit{lr} of $1 \cdot 10^{-3}$ and a \textit{wd} of $0.24$ to be optimal. We scale this \textit{lr} linearly by batch size ($0.24 \cdot 128 / 2048$) to obtain the used value of $6.25 \cdot 10^{-5}$.} by \cite{vitc}, using a learning rate of $6.25 \cdot 10^{-5}$ and a weight decay of $0.24$. Following \cite{vitc}, weight decay is not applied to the biases and any normalisation parameters. AdamW's $\beta$ parameters are set as $\beta_{1} = 0.9$ and $\beta_{2} = 0.999$. All experiments are run on a single NVIDIA RTX 6000 GPU.
\subsection{HEAR evaluation}
We use $18$ tasks from the HEAR 2021 Challenge \cite{hear}, derived from $15$ datasets, to evaluate the quality of the learned audio representations. HEAR includes two types of tasks, \textit{scene-based}, corresponding to classification (multi-class or multi-label) of an entire audio clip, and \textit{timestamp-based}, corresponding to sound event detection and or transcription over time. Scene-based tasks, following \citet{byol-s}, can be subdivided into three subcategories: speech, environmental sounds, and music. All datasets are downloaded\footnote{\url{https://zenodo.org/record/5887964}} at $48$ kHz and re-sampled to $16$ kHz to align with the sampling frequency of the AudioSet clips used during model pre-training. The re-sampling is implemented using the \texttt{librosa} Python library \cite{librosa}.

For each task, the embeddings are first extracted from the frozen pre-trained model. The timestamp-based tasks first require the input audio clips to be divided into fixed-size segments, such that embeddings can be extracted corresponding to specific timestamps. We use a segment size of $950$ ms with a hop size of $50$ ms for all timestamp-based tasks. The embeddings are then evaluated using the \texttt{hear-eval}\footnote{\url{https://github.com/hearbenchmark/hear-eval-kit}} toolkit, which trains a shallow MLP classifier on the frozen embeddings (linear evaluation). The MLP is trained for a maximum of $500$ epochs with the Adam optimizer \cite{adam}, implementing early stopping on the validation set, checking every $3$ epochs with a patience of $20$ (except for with DCASE 2016 Task $2$, which is checked every $10$ epochs). Model selection is performed over a choice of $8$ models each of which uses a different hyperparameter configuration, selecting the optimal model using the validation score. Variations in the number of hidden layers, learning rate, and weight initialization are considered. Full details can be found in the original HEAR publication \cite{hear}.

\section{Hyperparameter Search}\label{sec:hyps}

\begin{figure}[t]
\centering
\begin{subfigure}{.65\textwidth}
    \centering
    \includegraphics[width=\linewidth]{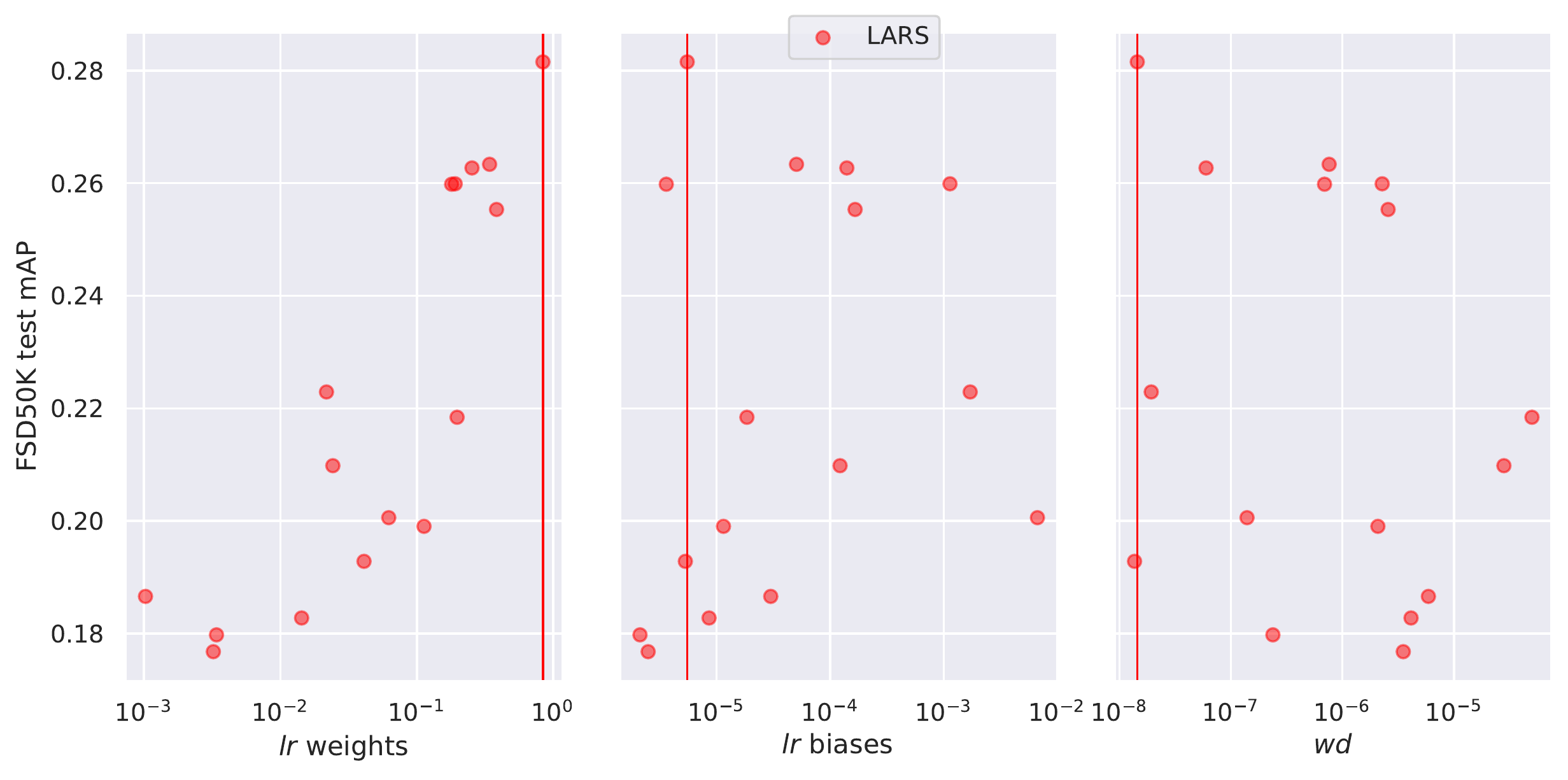}  
    \caption{
    LARS
    }
    \label{fig:AudioNTT_LARS}
\end{subfigure}
\begin{subfigure}{.65\textwidth}
    \centering
    \includegraphics[width=\linewidth]{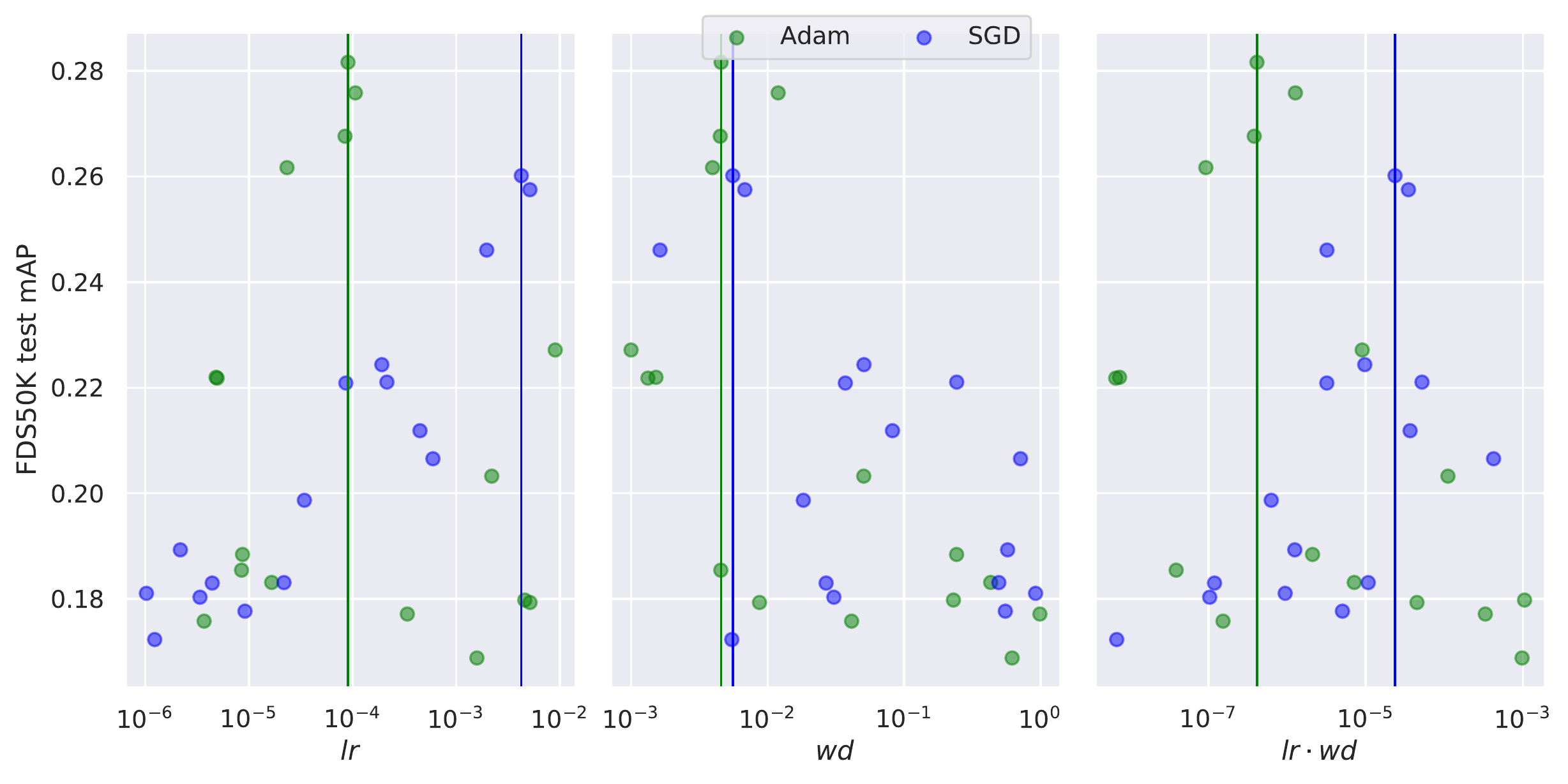}  
    \caption{
    Adam, SGD
    }
    \label{fig:AudioNTT_Adam_SGD}
\end{subfigure}
\caption{
AudioNTT optimizer hyperparameter sweep. Scatterplots are shown for all three optimizers: a) LARS (red), b) Adam (green), SGD (blue). FSD50K test mAP under linear evaluation is shown on the vertical axis and vertical bars correspond to the optimal values for each optimizer.
}
\end{figure}

\begin{figure}[t]
\centering
\begin{subfigure}{.65\textwidth}
    \centering
    \includegraphics[width=\linewidth]{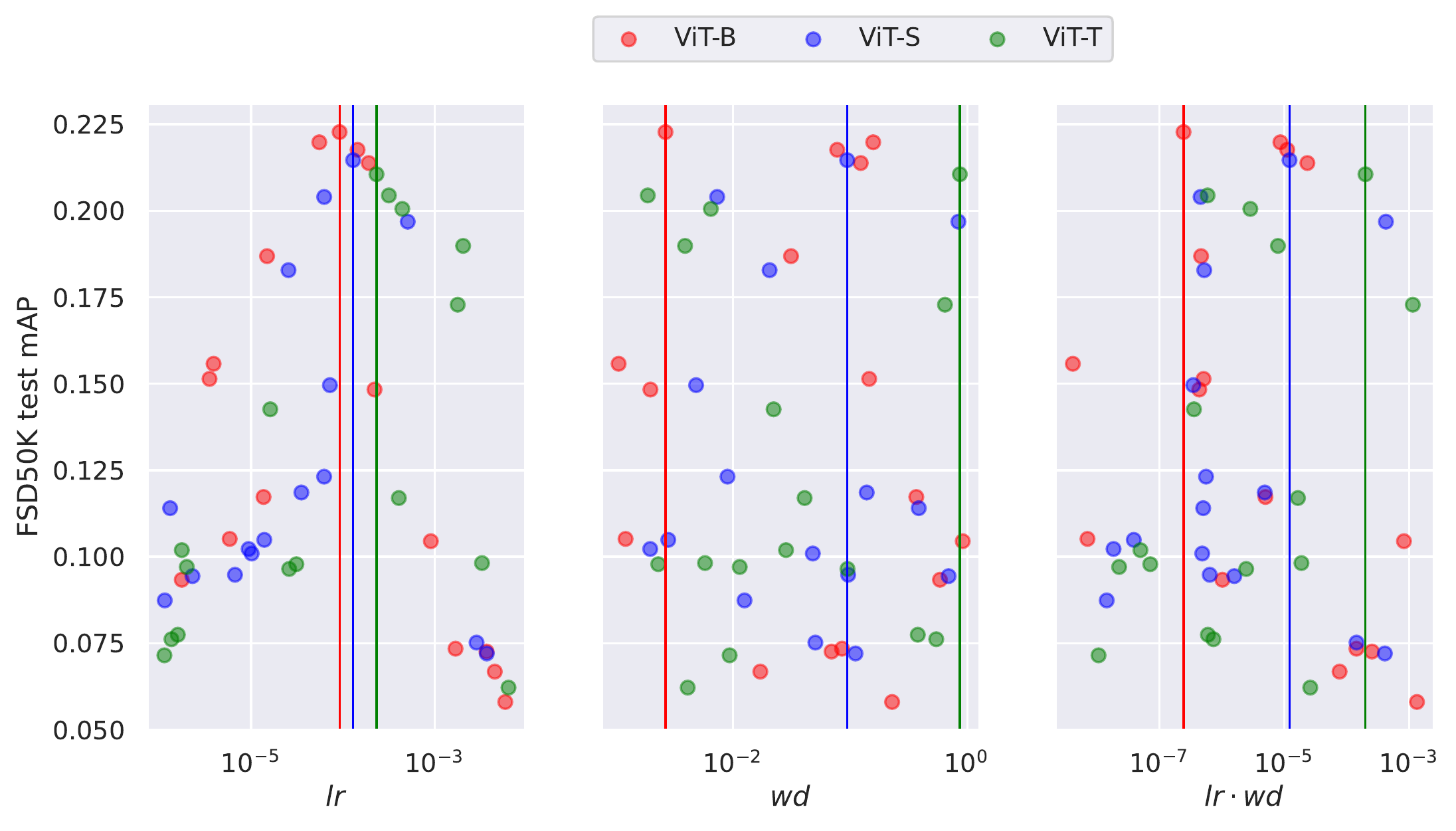}  
    \caption{ViT}
        \label{fig:ViT_AdamW}
\end{subfigure}
\begin{subfigure}{.65\textwidth}
    \centering
    \includegraphics[width=\linewidth]{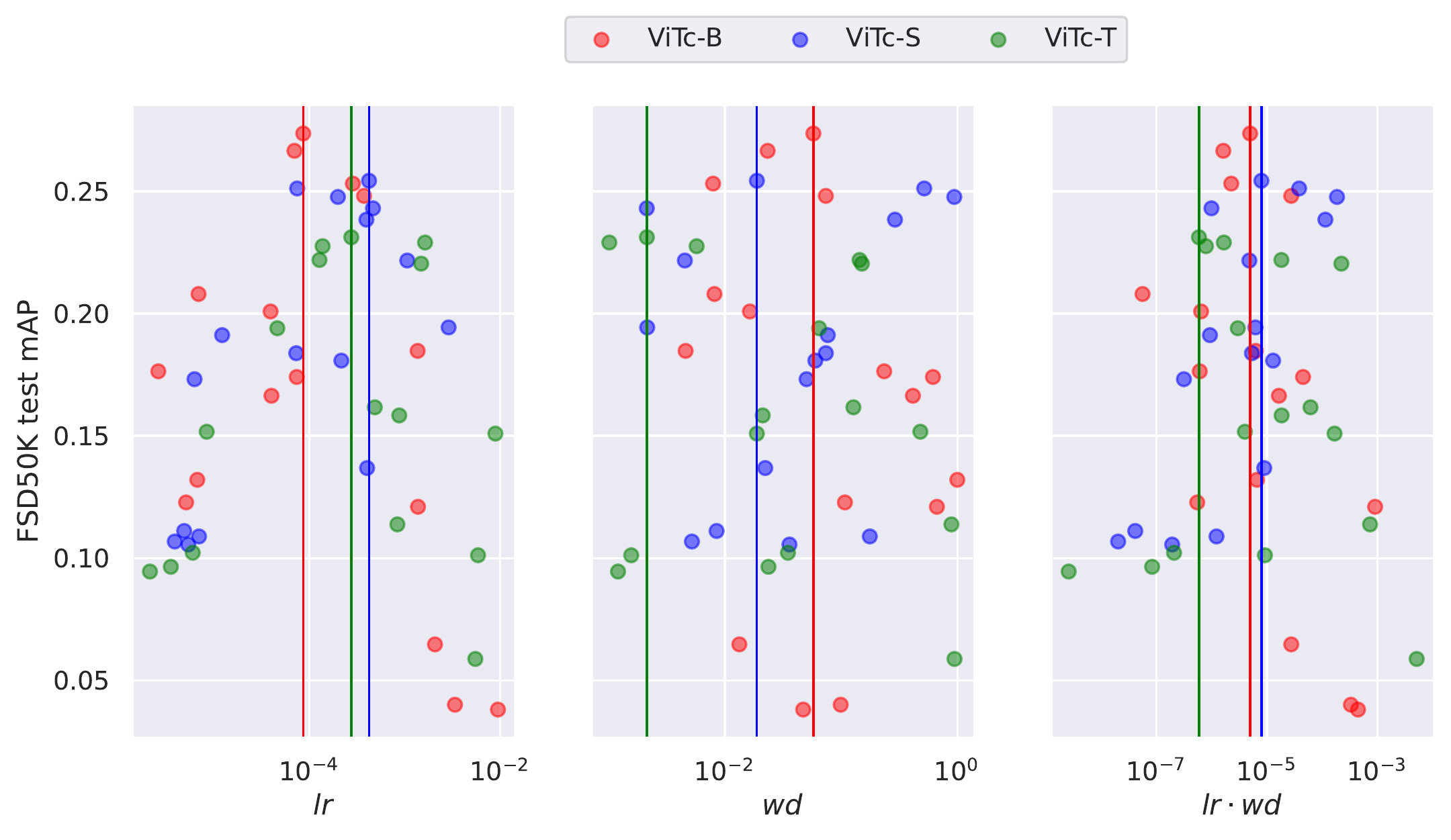}  
    \caption{ViT$_{C}$}
    \label{fig:ViTc_AdamW}
\end{subfigure}
\caption{
ViT / ViT$_{C}$ optimizer hyperparameter sweep. Scatterplots are shown for all three model sizes (-\textit{B} (red), -\textit{S} (blue), -\textit{T} (green)) for both the vanilla ViT (a) and ViT$_{C}$ (b) models, with FSD50K test mAP under linear evaluation shown on the vertical axis. Vertical bars correspond to the optimal values for each model.
}
\end{figure}

Before performing full ABT pre-training on AudioSet, hyperparameter sweeps are conducted over several important variables that we anticipate will most significantly affect optimization. We consider these to be the choice of optimizer, the optimizer learning rate (\textit{lr}) and weight decay (\textit{wd}).

For each hyperparameter configuration, ABT pre-training is performed on the FSD50K \cite{fsd50k} development subset for a maximum of $20$ epochs, corresponding to $\sim 6400$ training iterations (with a batch size of $128$). We measure performance through training a linear classifier\footnote{We train the linear classifier with the following set of hyperparameters: Adam optimizer, batch size $= 200$, \textit{lr} $= 1 \cdot 10^{-3}$, \textit{wd} $= 1 \cdot 10^{-8}$, Adam $\beta_{1} = 0.9$, Adam $\beta_{2} = 0.999$, Adam $\epsilon = 1 \cdot 10^{-8}$.} (linear evaluation protocol) on the frozen features of the FSD50K train subset, extracted from the pre-trained model, implementing early stopping on the FSD50K validation subset with a patience of $10$. The classifier is trained for a maximum of $100$ epochs. We report the model performance as the mean Average Precision (mAP) on the FSD50K evaluation subset. We anticipate this metric to be generally indicative of the quality of the learned audio representations for a given set of hyperparameter values. During linear evaluation on FSD50K, all audio clips are first randomly cropped to $96$ time frames, the same number as used during ABT pre-training. This is done to reduce the time to extract the embeddings from the pre-trained model. This explains the discrepancy between the FSD50K test scores (mAP) reported here and in Section \ref{sec:results}, where linear evaluation is also performed on FSD50K but with no cropping.

The hyperparameter sweeps are implemented using the \texttt{Optuna} framework \cite{optuna}, evaluating $16$ hyperparameter configurations (\textit{trials}) per sweep. Each set of hyperparameter values are sampled each trial using \texttt{Optuna}'s Tree-structured Parzen Estimator (TPE) sampler, and we evaluate model performance at the end of each of the $20$ pre-training epochs to allow pruning\footnote{We use \texttt{Optuna}'s Hyperband pruner to prune unpromising trials.} of unpromising trials at intermediate stages (before the maximum of $20$ epochs). All hyperparameter sweeps use the same audio preprocessing and ABT architectural defaults as in full AudioSet pre-training. That is, except for the projector output dimension, which has a slightly smaller value of $256$\footnote{We use a projector output dimension of 256 both in the hyperparameter sweeps and all ablation studies as initial experimentation suggests that this value is optimal. Extensive ablation studies, however, reveal that a larger value of $1048$ is preferable, and as such this value is used in full AudioSet pre-training.}. We measure the importance of individual hyperparameters (i.e. optimization sensitivity) for a given sweep using the fANOVA \cite{fanova} hyperparameter importance evaluation algorithm (available within the \texttt{Optuna} framework). fANOVA fits a random forest regression model to the scores of the completed (unpruned) trials for the trial hyperparameter configurations. The variance of the fitted model is then decomposed into additive components, each of which are associated with a specific hyperparameter. The fractional variance associated with a hyperparameter is taken as its importance.

\textbf{AudioNTT}\quad For the AudioNTT encoder, we consider pre-training with Adam, SGD and LARS. For Adam and SGD, each trial a (\textit{lr}, \textit{wd}) pair is sampled (on a log scale), considering \textit{lr} $= \{10^{-6}, 10^{-2}\}$ and \textit{wd} $= \{10^{-3}, 10^{0}\}$, with all other optimizer hyperparameters set to their \texttt{PyTorch} defaults\footnote{For SGD: Nesterov momentum $=$ False. For Adam: $\beta_{1} = 0.9$, $\beta_{2} = 0.999$, $\epsilon = 1 \cdot 10^{-8}$.}. Bias and batch normalisation parameters are excluded from weight decay. For LARS\footnote{The other LARS hyperparameters (see \cite{lars}) are set as: momentum $m = 0.9$, LARS coefficient $\eta = 1 \cdot 10^{-3}$}, we consider separate \textit{lr}s for the weights and for the biases, considering \textit{lr weights} $ = \{10^{-3}, 10^{0}\}$, \textit{lr biases} $= \{10^{-6}, 10^{-2}\}$, and \textit{wd} $= \{10^{-8}, 10^{-4}\}$. A (\textit{lr weights}, \textit{lr biases}, \textit{wd}) triplet is sampled (on a log scale) each trial. For all three optimizers a total of $16$ trials are evaluated. Figures \ref{fig:AudioNTT_LARS} (LARS) and \ref{fig:AudioNTT_Adam_SGD} (Adam, SGD) show scatterplots for the FSD50K test scores (mAP) for the models trained with the three optimizers. We observe that LARS and Adam both attain higher optimal performance ($\sim 0.28$) than SGD ($\sim 0.26$). LARS is also less sensitive to the \textit{wd} value than Adam and SGD, and as such LARS is used as the default optimizer for pre-training with the AudioNTT encoder\footnote{The choice of LARS as the default optimizer (with a convolutional encoder) is further motivated by that LARS is used in the original Barlow Twins publication \cite{barlow_twins}}. From Fig.\ref{fig:AudioNTT_LARS}, we see that LARS tends to prefer a larger value for \textit{lr weights} $\sim O(10^{0})$, showing less sensitivity to \textit{lr biases} and \textit{wd}, with importance values of $0.81$, $0.10$, and $0.09$, respectively. The optimal trial, which achieves a score of $0.282$, uses (\textit{lr weights}, \textit{lr biases}, \textit{wd}) values of $(0.84, 5.5 \cdot 10^{-6}, 1 \cdot 10^{-8})$. We find in general that (\textit{lr weights}, \textit{lr biases}, \textit{wd}) values of $(0.4, 4.8 \cdot 10^{-3}, 1 \cdot 10^{-6})$ are effective in general, and as such these values are used with the LARS optimizer by default.

\textbf{ViT$\mathbf{_{C}}$}\quad We perform hyperparameter sweeps for both the ViT$_{C}$ \cite{vitc} and vanilla ViT \cite{vit} encoders, considering only the AdamW\footnote{\cite{vitc} find that the ViT$_{C}$ models are also stable when trained with SGD. However, initial experimentation reveals that training ViT$_{C}$ encoders with SGD (with ABT pre-training) is unstable, with the loss frequently going to NaN.} optimizer. We use a patch size of $16 \times 16$. Each trial, a (\textit{lr}, \textit{wd}) pairs is sampled (on a log scale), considering \textit{lr} $= \{10^{-6}, 10^{-2}\}$ and \textit{wd} $= \{10^{-3}, 10^{0}\}$, with all other hyperparameters set to their \texttt{PyTorch} defaults\footnote{For AdamW: $\beta_{1} = 0.9$, $\beta_{2} = 0.999$, $\epsilon = 1 \cdot 10^{-8}$}. Bias and batch normalisation parameters are excluded from weight decay. Hyperparameter sweeps are conducted for all ViT and ViT$_{C}$ model sizes (-\textit{B}, -\textit{S}, -\textit{T}), evaluating a total of $16$ trials for each model. Figures \ref{fig:ViT_AdamW} (ViT) and \ref{fig:ViTc_AdamW} (ViT$_{C}$) show scatterplots for the FSD50K test scores (mAP) for the ViT and ViT$_{C}$ models. We observe that for all ViT and ViT$_{C}$ models, a \textit{lr} of $\sim 1 \cdot 10^{-4}$ is optimal, although the smaller models (-\textit{S}, -\textit{T}) tend to prefer a slightly larger value. The models show less sensitivity to the \textit{wd} value. In general, a \textit{lr}/\textit{wd} of $1 \cdot 10^{-4}$/$0.06$ is effective for all ViT and ViT$_{C}$ models, and as such these values are used by default during the ablation studies (Appendix \ref{sec:ablat}). However, we find that the tuned values used by \citet{vitc} for the ViT$_{C}$ models, a \textit{lr}/\textit{wd} of $6.25 \cdot 10^{-5}$/$0.24$, lead to better model performance\footnote{Model performance as measured by tracking FSD50K linear evaluation score every $5$ epochs during training.} when pre-training on the full AudioSet unbalanced train subset. We anticipate that this is because a slightly lower \textit{lr} is preferred for a significantly longer training schedule\footnote{$100$ epochs on AudioSet unbalanced train segments corresponds to $\sim 250$ $\times$ training iterations as with $20$ epochs on the FSD50K development subset.}. We also note three further salient points: 1) The ViT$_{C}$ models reach a higher optimal performance than their corresponding vanilla ViT models for all model sizes (-\textit{B}: $0.27 \text{ vs } 0.22$,-\textit{S}: $0.25 \text{ vs } 0.21$, -\textit{T}: $0.23 \text{ vs } 0.21$), 2) The ViT$_{C}$ models show less sensitivity to the exact \textit{lr} value used than the ViT models, with wider peaks for all three model sizes, supporting the stability claims made by \cite{vitc}, 3) Both ViT and ViT$_{C}$ optimal performance scales with model size. Point 1) motivates only using the ViT$_{C}$, and not vanilla ViT, encoders for full AudioSet ABT pre-training, since the ViT$_{C}$ encoders significantly outperform them and full AudioSet pre-training with a Transformer encoder takes several days to complete\footnote{ABT pre-training on the unbalanced train subset of AudioSet with a ViT${C}$-\textit{B} on a single NVIDIA RTX 6000 GPU, using mixed precision, takes approximately 120 hours.}.

\section{Ablation Studies}\label{sec:ablat}

\begin{figure}[h]
\centering
\includegraphics[width=0.6\textwidth]{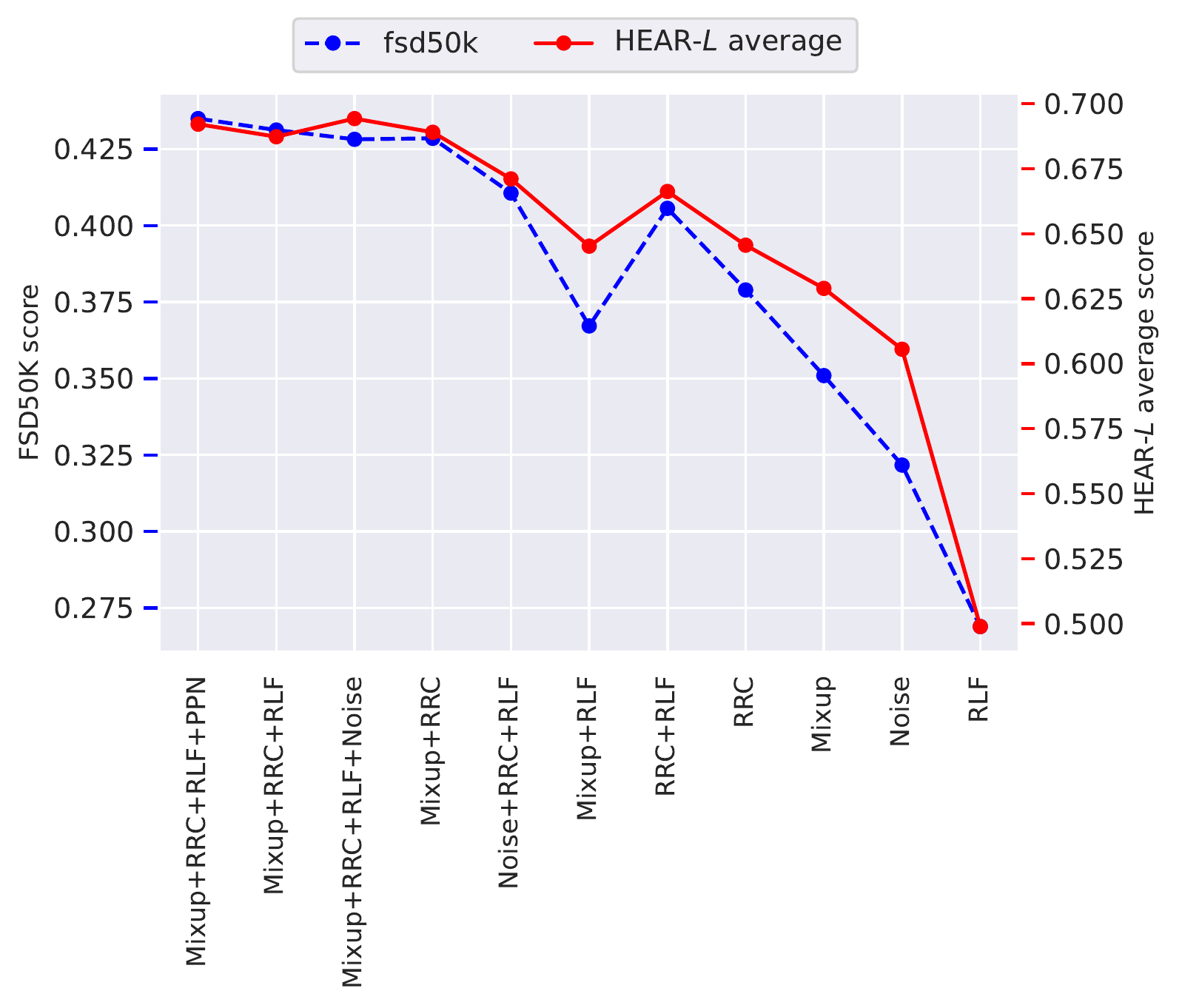}
\caption{
We compare the effect of pre-training with different combinations of the components of the Audio Barlow Twins audio augmentation (AA) module. Results are shown both evaluated on FSD50K (blue) and the average score on the $5$ HEAR-\textit{L} tasks (red).
}
\label{fig:ablations_data_aug}
\end{figure}

We perform extensive ablation studies to investigate the contributions of each of the different components of the ABT learning framework. For all ablation studies (except for those in Appendix \ref{sec:ablat_masking}, which use a ViT$_{C}$-\textit{B} encoder), we perform ABT pre-training for $100$ epochs with the AudioNTT encoder on the FSD50K development subset, which corresponds to $\sim 32$k training iterations (with a batch size of $128$). Asides from the training duration and training dataset, all ablation studies use the same experimental set-up as used for full 
AudioSet pre-training, except for the projector output dimension, which is set by default to $256$. For all ablations considered, we evaluate model performance through linear evaluation, using the \texttt{hear-eval} toolkit, on a lightweight version of the HEAR Challenge, which we term HEAR-\textit{L}. HEAR-\textit{L} consists of five HEAR tasks covering all three of the scene-based task subcategories: CREMA-D (speech), LibriCount 
(speech), FSD50K (environmental sound), ESC-50 (environmental sound), and GTZAN Genre (music).

\subsection{Audio Augmentations}\label{sec:ablat_aa}

We consider using different combinations of the components of the audio augmentation (AA) module, which by default consists of Mixup, Random Resize Crop (RRC), and Random Linear Fader (RLF). We further consider two different variations, namely Pre-Post-Norm (\textit{PPN}) and \textit{Noise}. PPN refers to removal of the normalisation block, which standardises input spectrograms by the dataset mean and standard deviation, and replacing it with the Pre- and Post-Normalisation blocks proposed by \citet{byol-a} in BYOL-A. Specifically, the pre-normalisation block normalises input spectrograms by batch (and not dataset) statistics, and the post-normalisation block does the same, but after the application of the audio augmentations (Mixup, RRC, RLF). \citet{byol-a} argue the post-normalisation corrects the statistical drifts caused by the applied augmentations. Noise refers to the addition of random noise\footnote{We sample the noise from a Gaussian distribution $\mathcal{N}(0, \lambda)$, where $\lambda \sim U(0, \alpha)$, $\alpha = 0.2$.} to an incoming spectrogram. We implement this for direct comparison with Mixup, which interpolates the incoming spectrogram with a natural background signal randomly sampled from the training dataset.

From Figure \ref{fig:ablations_data_aug} we note four salient points:

\textbf{1. Strong audio augmentations are essential to learn high-quality representations}\newline
When all the augmentations are removed from the baseline except RLF (remove RRC and Mixup), Audio Barlow Twins performance drops significantly, by $19$ points from $69\%$ to $50\%$ average on the HEAR-\textit{L} tasks.

\textbf{2. Mixing with natural background signals is more effective than with noise}\newline
Mixup improves $13$ points from the RLF average on HEAR-\textit{L} to $63\%$, whereas addition of Gaussian noise results in a smaller improvement of only $10\%$. Further, using Noise as well as Mixup+RRC+RLF leads to no significant additional performance improvements.\newline

\textbf{3. RRC is the most effective audio augmentation}
RRC, which approximates pitch shift and time stretch, attains the highest performance when any of the audio augmentations are applied alone, achieving a HEAR-\textit{L} average score of $65\%$ (compared with $63\%$ for Mixup, $60\%$ for Noise, and $50\%$ for RLF). RLF is by far the least effective augmentation. These findings are consistent with previous results found by \cite{byol-av2}.\newline

\textbf{4. PPN shows minimal improvement over dataset normalisation}
Mixup+RRC+RLF+PPN results in almost identical model performance as Mixup+RRC+RLF (with normalisation by dataset statistics), both having a HEAR-\textit{L} average score of $\sim 69\%$.\newline

\subsection{Learning Framework}\label{sec:ablat_lf}

\begin{figure}[ht]
\begin{subfigure}{.475\linewidth}
  \includegraphics[width=\linewidth]{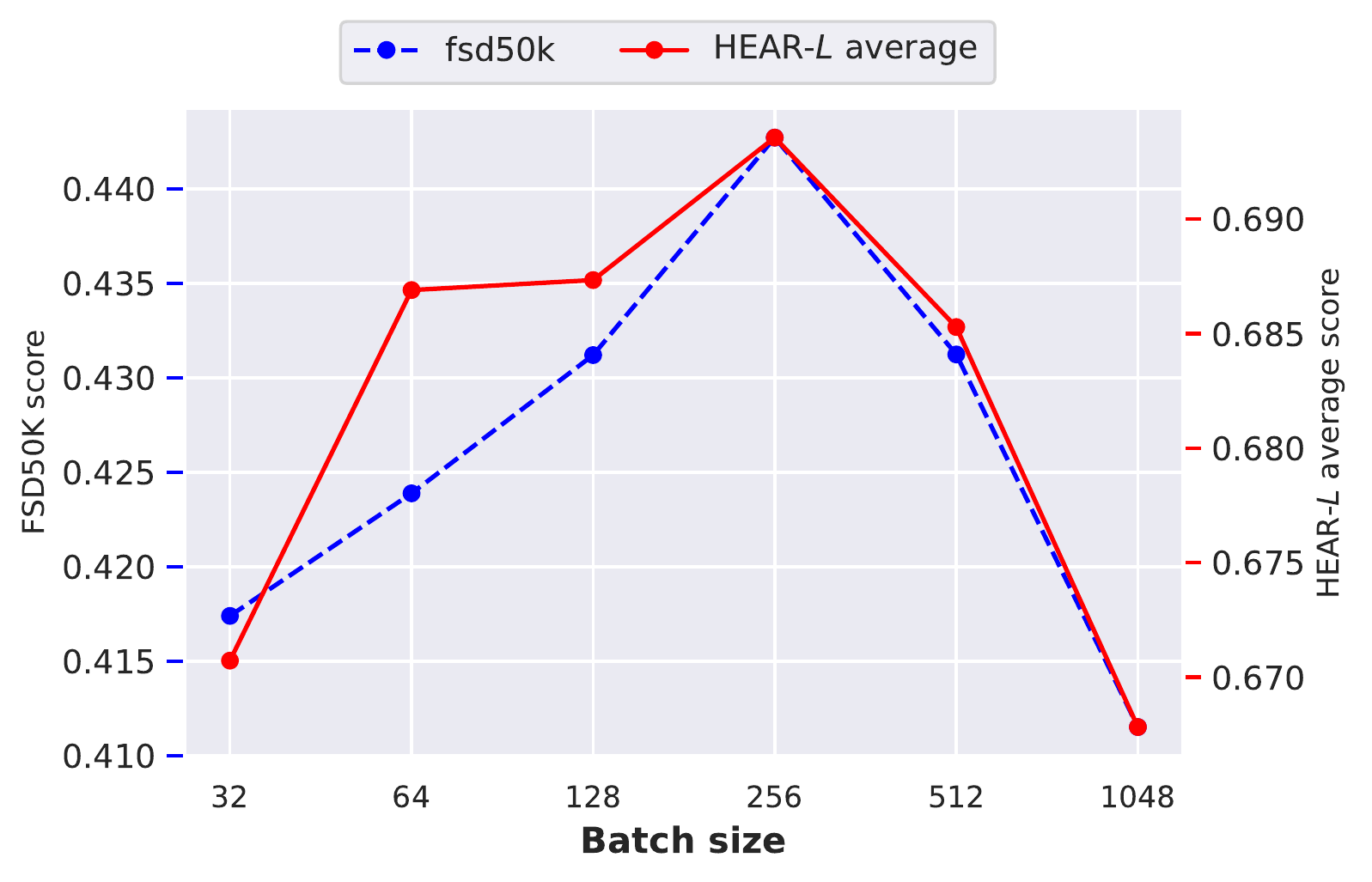}
  \caption{}
  \label{fig:ablations_batch_size}
\end{subfigure}\hfill
\begin{subfigure}{.475\linewidth}
  \includegraphics[width=\linewidth]{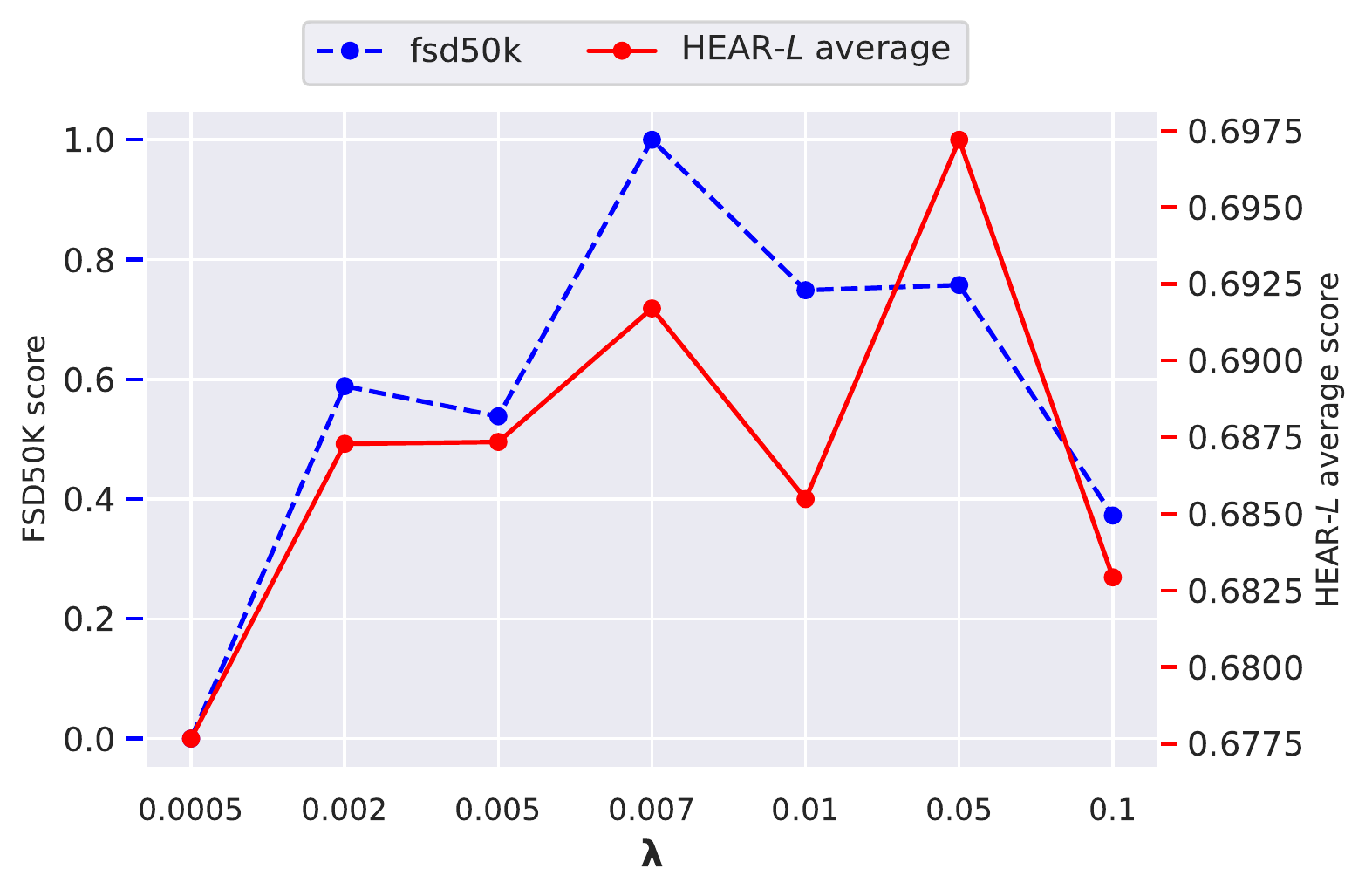}
  \caption{}
  \label{fig:ablations_lambda}
\end{subfigure}
\medskip
\begin{subfigure}{.475\linewidth}
  \includegraphics[width=\linewidth]{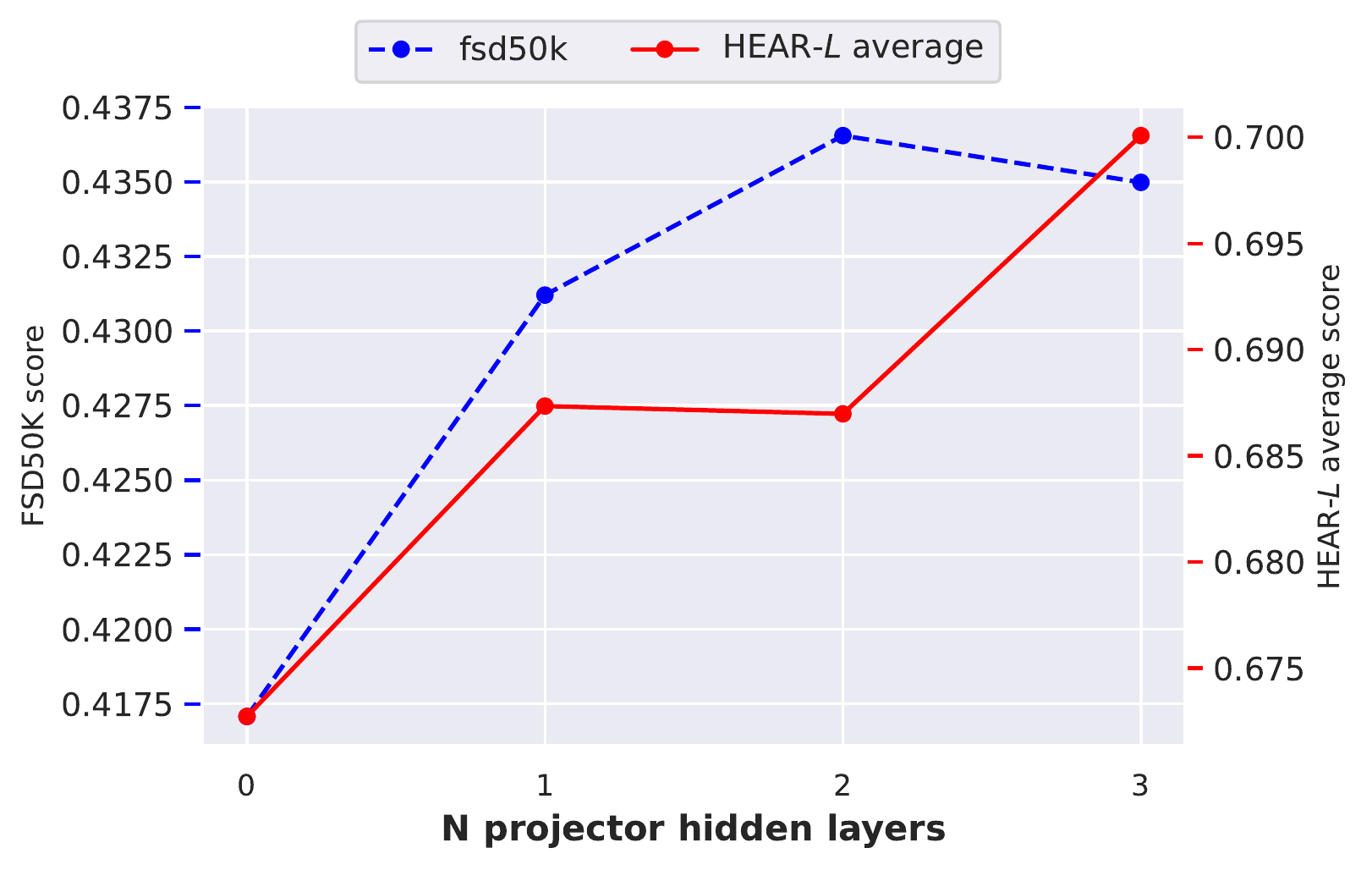}
  \caption{}
  \label{fig:ablations_n_hidden_layers}
\end{subfigure}\hfill 
\begin{subfigure}{.475\linewidth}
  \includegraphics[width=\linewidth]{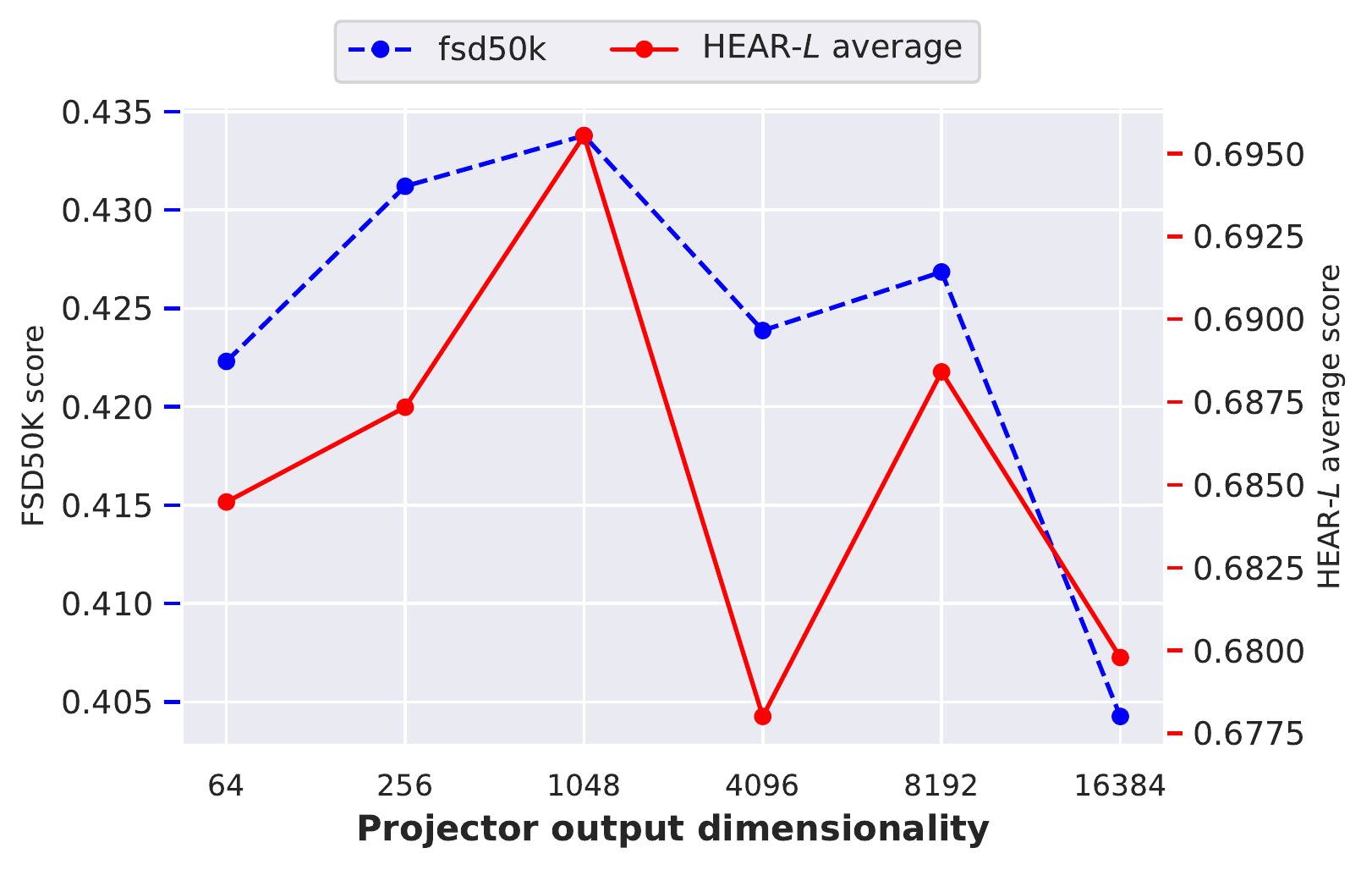}
  \caption{}
  \label{fig:ablations_out_dim}
\end{subfigure}
\caption{
Ablation studies for \textbf{(a)} batch size, \textbf{(b)} Barlow Twins objective hyperparameter $\lambda$, \textbf{(c)} projector depth, and \textbf{(d)} projector output dimensionality. Results are shown both evaluated on FSD50K (red) and the average score on the $5$ HEAR-\textit{L} tasks (blue).
}
\label{fig:ablations_bt_architecture}
\end{figure}

\begin{figure}[ht]
\centering
\includegraphics[width=0.6\textwidth]{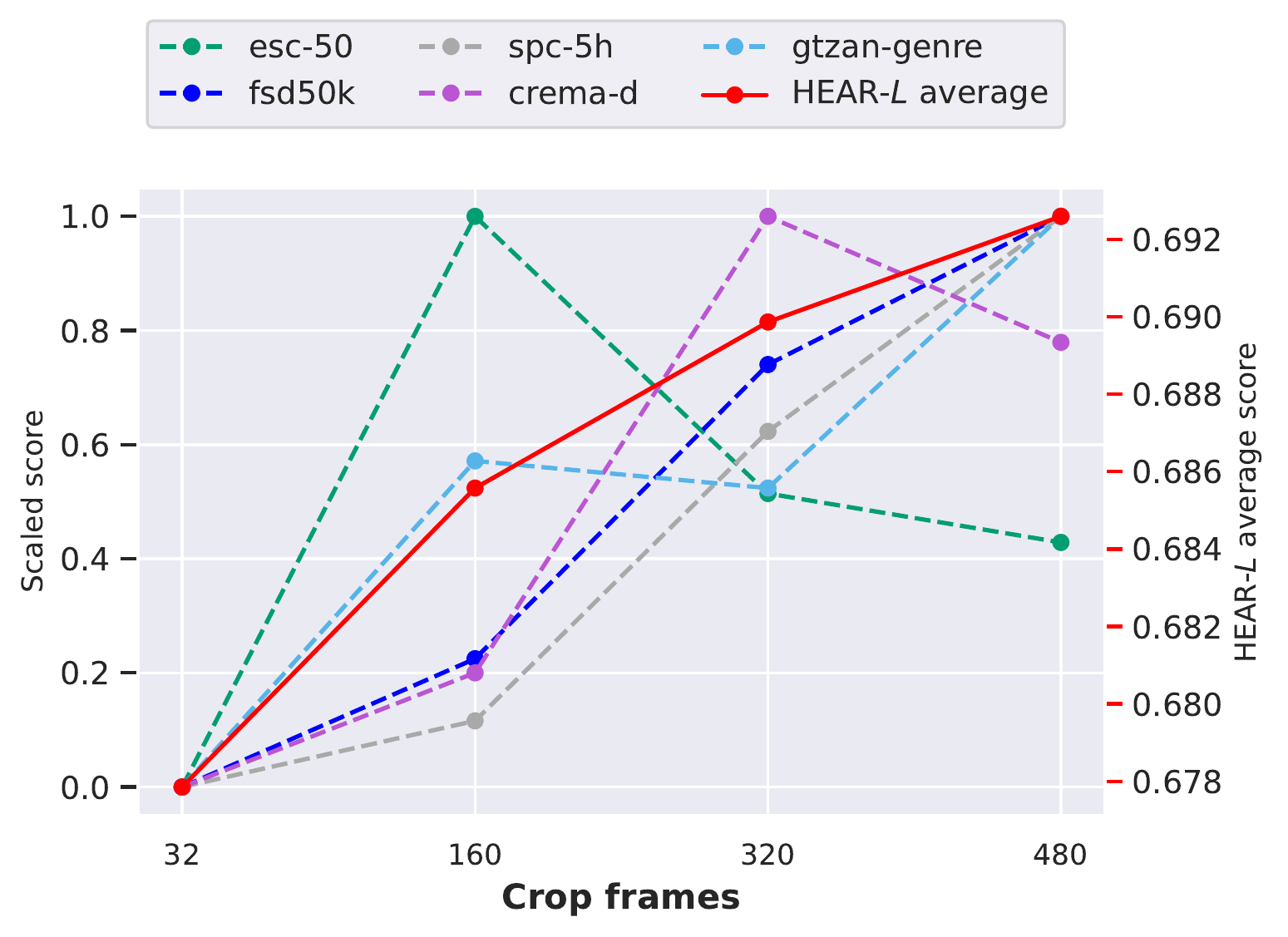}
\caption{
We compare the effect of pre-training with a different length of the input audio used during pre-training, considering cropping the input spectrograms to $32$, $160$, $320$, and $480$ frames. Results are shown evaluated on the individual HEAR-\textit{L} tasks (ESC-50 (green), Speech Commands $5$h (grey), GTZAN Genre (light blue), FSD50K (dark blue), CREMA-D (purple), as well as the HEAR-\textit{L} average (red). Since the scores on the individual HEAR-\textit{L} tasks have different scales, we show the MinMax scaled scores, where the best performing input length for each task is set to $1$ and the worst performing to $0$.
}
\label{fig:ablations_input_length}
\end{figure}

\textbf{Batch Size}\quad Figure \ref{fig:ablations_batch_size} shows the sensitivity of Audio Barlow Twins to batch size. The size of the batch is expected to influence training, and therefore downstream performance, since batch dynamics contribute significantly to the Barlow Twins objective (Eqn.\ref{Barlow Twins loss}) through the empirical cross-correlation matrix, which is computed across the batch embeddings (Eqn.\ref{cross-correlation matrix}: $C_{ij} = \sum_{b=1}^{B} \hat{Z}_{\theta, i}^{b} \hat{Z^{\prime}}_{\theta, j}^{b}$). We observe that ABT exhibits reasonable sensitivity to batch size, with strongest performance with a medium-sized batch containing $64-512$ samples. This is contrary to methods in CV such as SimCLR \cite{simclr}, which prefer much larger batch sizes (SimCLR requires a batch size of at least $1048$ for strong performance). However, we note that all models are trained with the learning rates (\textit{lr weights} and \textit{lr biases}) tuned with a batch size of $128$, applying linear scaling: $\text{\it{lr}} = \text{\it{lr}}_{128} \times \text{BatchSize} / 128$. Re-tuning LARS learning rates for each batch size is beyond the scope of this project. We are unable to consider batch sizes above $1048$ due to GPU memory restrictions.

\textbf{$\bf{\lambda}$, Projector Depth, Projector Output Dimensionality} \quad Figures \ref{fig:ablations_lambda}, \ref{fig:ablations_n_hidden_layers}, and \ref{fig:ablations_out_dim} show the variation of model performance with the Barlow Twins objective hyperparameter $\lambda$, the number of hidden dimensions of the projector network, and the dimensionality of the embeddings (over which the Barlow Twins objective is calculated). We observe that ABT shows minimal sensitivity to the exact value of $\lambda$, as found by \citet{barlow_twins} in the original Barlow Twins publication, although a value in the approximate range $0.002 < \lambda < 0.05$ is preferred, allowing for both the invariance and redundancy reduction terms of the Barlow Twins objective (Eqn.\ref{Barlow Twins loss}) to contribute. We further observe that a deeper projector is preferred, although performance does not significantly rise above a depth of $2$ ($1$ hidden layer). Contrary to the observations of \citet{barlow_twins}, we don’t find that model performance continues to improve as projector output dimensionality grows, with saturation at an output size of $1048$, and considerable performance degradation observed with a dimensionality of $16,384$.

\textbf{Input Audio Duration}\quad We consider variations in the length of the input audio used during ABT pre-training. Specifically, we consider cropping the input spectrograms to $32$, $160$, $320$, and $480$ frames, which correspond to $\sim$ $320$ms, $1.6$s, $3.2$s, and $4.8$s of audio, respectively. We choose not to show the results with the default $96$ crop frames as all hyperparameters have been tuned using this value, and as a result it is not considered to be a fair comparison. As shown in Figure \ref{fig:ablations_input_length}, almost all HEAR-\textit{L} tasks benefit from a longer training window, with the exception of ESC-50. For speech tasks, such as Speech Commands or CREMA-D, this is intuitive, since an element of speech may last several seconds, and as such using only a short segment of under one second in pre-training can result in sounds that don’t retain the original semantic content of the clip (e.g. a word may be cropped to only a syllable or single character). However, ESC-50 clearly seems to benefit from a shorter input duration, showing optimal performance with $160$ input frames. The environmental sound dataset ESC-50 contains many sounds categories which consist of short, sharp noises, such as the categories mouse click and door knock, and as such using a short segment during pre-training may better align with the actual sound duration of this dataset. The optimal input duration during pre-training therefore appears to be dependent on the downstream dataset being evaluated on, although there exists a general trend that longer clips are beneficial.

\subsection{View Masking}\label{sec:ablat_masking}

\begin{figure}[t]
\begin{subfigure}{.475\linewidth}
  \includegraphics[width=\linewidth]{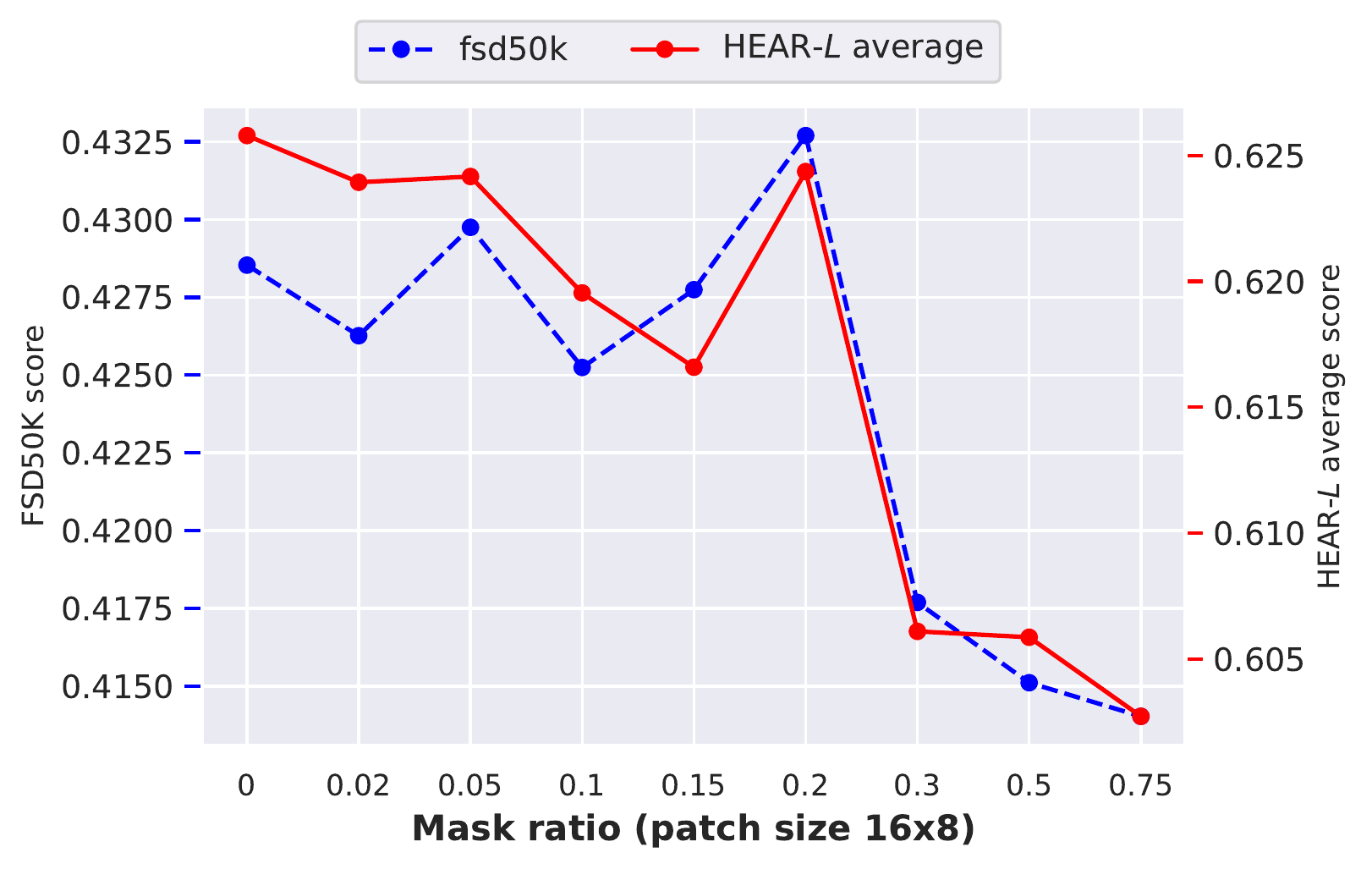}
  \caption{}
  \label{fig:vit_ablations_masking_16x8}
\end{subfigure}\hfill
\begin{subfigure}{.475\linewidth}
  \includegraphics[width=\linewidth]{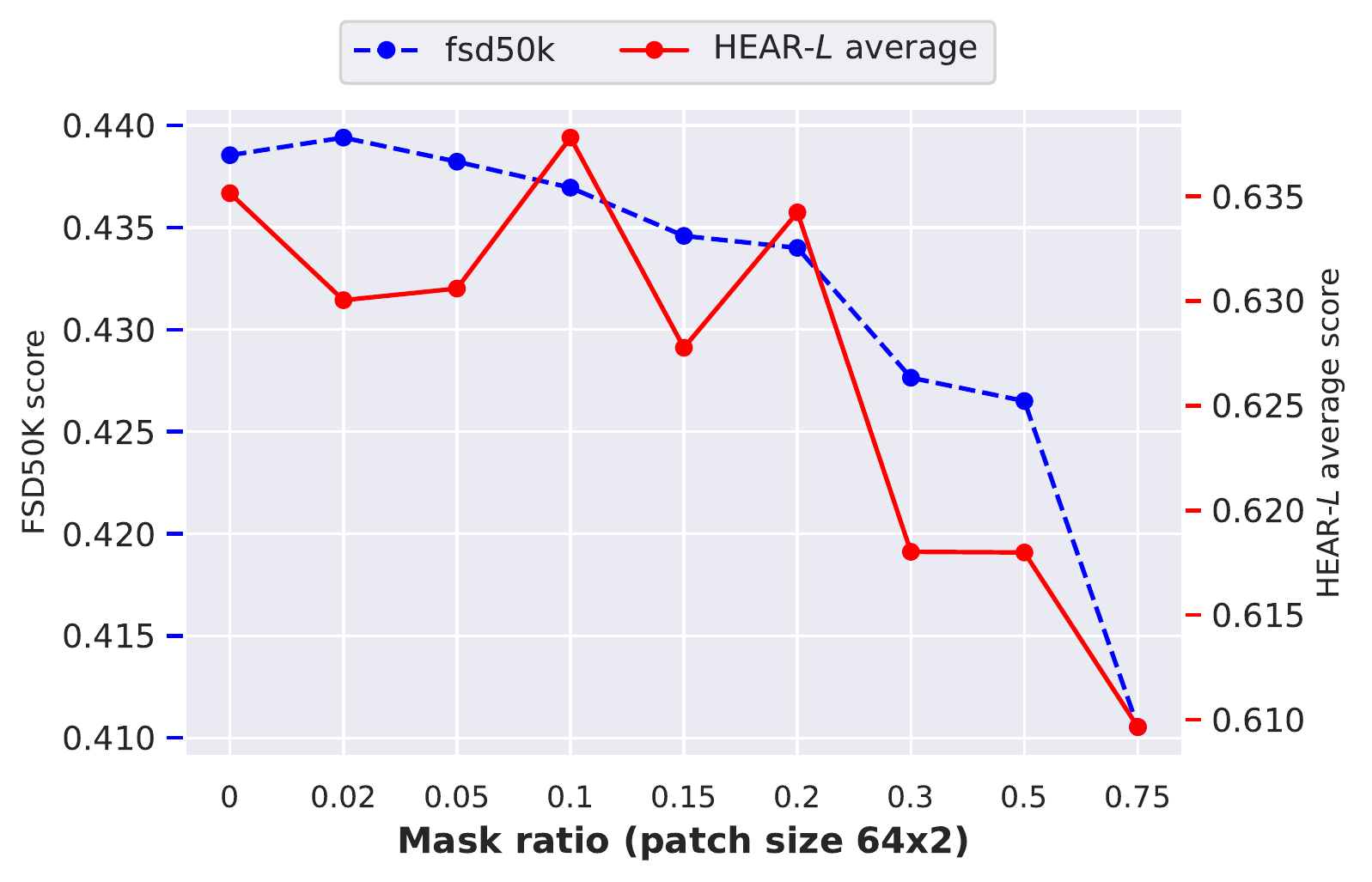}
  \caption{}
  \label{fig:vit_ablations_masking_64x2}
\end{subfigure}
\caption{
We consider the effect of partial view masking of one of the two spectrogram views, inspired by recent work by \cite{msn}. We pre-train with different masking ratios, considering masking with both \textbf{(a)} $16 \times 8$ patches and \textbf{(b)} $64 \times 2$ patches. Results are shown both evaluated on FSD50K (blue) and the average score on the $5$ HEAR-\textit{L} tasks (red).
}
\label{fig:vit_ablations_masking}
\end{figure}

In their recent work \textit{Masked Siamese Networks} (MSN), \citet{msn} randomly drop a subset of the patches of one of the two image views before being processed by a Siamese Network architecture using a ViT encoder, matching in feature space the masked view with the unmasked view and thereby performing “implicit denoising” \cite{msn} at the representation level. MSN achieves SOTA results whilst simultaneously reducing computational and memory requirements, since the masked patches can be dropped before input into the ViT encoder. It is therefore of great interest to see whether adapting this approach to the audio domain can be beneficial in the pursuit of universal audio representations. Similarly to MSN, we consider adding the step of randomly masking patches from one of the two spectrogram views before input into the ViT$_{C}$ encoder to ABT's augmentation module. We implement random patch masking using the algorithm proposed by \citet{mae}, where the list of extracted patches is randomly shuffled, and the last $M$ patches from the list are removed, where $M = rN$ (rounded to the nearest integer), with $r$ being the masking ratio and $N$ the initial number of patches. We consider partial view masking with both a patch size of $16 \times 8$ and $64 \times 2$, which both correspond to a total of $N = 48$ patches (with $64 \times 96$ spectrogram inputs).

Disappointingly, as shown in Figures \ref{fig:vit_ablations_masking_16x8} and \ref{fig:vit_ablations_masking_64x2}, partial view masking seems to harm the quality of the learned audio representations, with a clear trend that over a minimum threshold for $r$ ($r \sim 0.2$, corresponding to $M = 10$ masked patches), model performance is significantly reduced. Below this threshold we generally see slight degradation in model performance, although minimal variation (expected as only very few patches have been masked). We anticipate that masking a large number of spectrogram patches may fundamentally change the semantic content of the audio clip, such that matching the representations of the masked and unmasked views encourages the model to embed together audio samples in representation space which no longer share the same semantic content, thereby damaging the quality of the learned representations. This is different to in CV, where strong masking doesn't visually appear to change the overall semantic content contained within an image (e.g. a heavily masked picture of a dog is still recognisable as a dog).

We additionally consider whether, instead of using a fixed masking ratio, slowly increasing the masking ratio during pre-training leads to improved representation quality. Starting the masking ratio at $0$, we increase it to a value $\beta$ at epoch $100$ following a sinusoidal schedule with a warm up period of $10$ epochs. However, initial experimentation with $\beta = 0.3$ suggests that this also results in a degradation of model performance, although extensive analysis has not been performed.

\end{appendices}
\end{document}